\documentstyle[12pt]{article}

\renewcommand{\thefootnote}{\fnsymbol{footnote}}

\def\NPB#1#2#3{Nucl. Phys. B{#1} (19#2) #3}
\def\PLB#1#2#3{Phys. Lett. B{#1} (19#2) #3}

\def\ov{\overline}
\def\s2{\frac{1}{\sqrt2}}
\def\os2{\frac{1}{2 \sqrt2}}

\def\oh{\frac{1}{2}}
\def\beq{\begin{equation}}
\def\eeq{\end{equation}}
\def\beqa{\begin{eqnarray}}
\def\eeqa{\end{eqnarray}}
\def\tr{{\rm tr \,}}
\def\Tr{{\rm Tr \,}}
\def\IF{\relax{\rm I\kern-.18em F}}
\def\II{\relax{\rm I\kern-.18em I}}
\def\IZ{\relax\ifmmode\hbox{\ss Z\kern-.4em Z}\else{\ss Z\kern-.4em Z}\fi}
\def\IP{\relax{\rm I\kern-.18em P}}
\def\inbar{\vrule height1.5ex width.4pt depth0pt}
\def\IC{\relax\hbox{\kern.25em$\inbar\kern-.3em{\rm C}$}}
\def\nh{\relax\ifmmode {n_H}\else $n_H$\fi }
\def\nt{\relax\ifmmode {n_T}\else $n_T$\fi }
\def\nv{\relax\ifmmode {n_V}\else $n_V$\fi }
\def\cp#1{\relax\ifmmode {\IP\kern-2pt{}_{#1}}\else $\IP\kern-2pt{}_{#1}$\fi}

\def\P1{\relax\ifmmode{\IP_1}\else $\IP_1$\fi}
\def\ds{\displaystyle}

\topmargin
-2cm
\textwidth
15cm
\textheight
24cm
\oddsidemargin
1cm
\evensidemargin
1cm
\arraycolsep        
2pt          
\begin{document}

\makeatletter
\@addtoreset{equation}{section}
\makeatother
\renewcommand{\theequation}{\thesection.\arabic{equation}}
\pagestyle{empty}
\rightline{ FTUAM-96/27, UTTG-09-96}
\rightline{\tt hep-th/9607121}
\begin{center}
\LARGE{
New Branches of String
Compactifications and\\ their F-Theory Duals}\\[5mm]
\large{G.~Aldazabal\footnote{Permanent Institutions: 
CNEA, Centro At\'omico
Bariloche, 8400 S.C. de Bariloche, and CONICET, Argentina.}$^{1}$,
A. Font\footnote{On sabbatical leave from
Departamento de F\'{\i}sica, Facultad de Ciencias,
Universidad Central de Venezuela. Work supported in part by the
N.S.F. grant PHY9511632 and the Robert A. Welch Foundation.}$^{2}$,
L.~E.~Ib\'a\~nez$^1$
and A.~M.~Uranga$^1$}\\[0.1in]
\begin{tabular}{cc}
\small $^1$ Departamento de F\'{\i}sica Te\'orica, &
\small $^2$ Theory Group, Department of Physics,\\[-3mm]
\small Universidad Aut\'onoma de Madrid, &
\small The University of Texas,\\[-3mm]
\small Cantoblanco, 28049 Madrid, Spain. &
\small Austin, TX 78712, USA.
\end{tabular}
\\[0.5in]
\small{\bf Abstract} \\[0.2in]
\end{center}

\begin{center}
\begin{minipage}[h]{5.5in}
We study heterotic $E_8\times E_8$  models that are 
dual to compactifications of F-theory and type IIA string on 
certain classes of elliptically fibered
Calabi-Yau manifolds.  Different choices for the specific torus 
in the fibration have heterotic duals that are most easily understood 
in terms of $E_8\times E_8$  models with gauge backgrounds of type
$H\times U(1)^{8-d}$, where $H$ is a non-Abelian factor. 
The case with $d=8$  corresponds to
the well  known  $E_8\times E_8$ compactifications with
non-Abelian instanton backgrounds $(k_1,k_2)$  whose F-theory duals
are built  through compactifications on fibrations of
the torus $\IP_2^{(1,2,3)}[6]$ over $\IF_n$. The new cases with 
$d < 8$ correspond to other choices for the
elliptic fiber over the same base and yield unbroken $U(1)$'s, some of
which are anomalous and acquire a mass by swallowing zero modes
of the antisymmetric $B_{MN}$ field. 
We also study transitions to models with no tensor multiplets in $D=6$ 
and find evidence of $E_d$ instanton dynamics.
We also consider the possibility of conifold transitions among 
spaces with different realization of the elliptic fiber.

\end{minipage}
\end{center}
\newpage

\setcounter{page}{1}
\pagestyle{plain}
\renewcommand{\thefootnote}{\arabic{footnote}}
\setcounter{footnote}{0}

\section{Introduction}

\bigskip

The existence of strong-weak coupling dualities \cite{sdual}
in string theory seems to be firmly established by now.
Many different strings on different vacua which were
previously thought to be independent turn out to be 
connected in some manner  by different string dualities.
More specifically, 
the evidence supporting the idea \cite{kv, fhsv}  
of a strong-weak coupling duality between type IIA and heterotic strings 
has increased with the new insights provided from the perspective
of F-theory \cite{vafa, mv, sen}. In this article we wish to explore
new branches of $K3$ and $K3\times T^2$ 
heterotic compactifications and explain
how they are related to F-theory and type IIA compactifications.

Our basic motivation is the observation of \cite{afiq} that in many cases
type II candidates to heterotic duals appear to be organized into chains, 
corresponding to sequential Higgsing in the heterotic side,
following a very  precise pattern. Duality requires the occurrence of
transitions among the Calabi-Yau (CY) spaces in the type II side as well 
as enhancing of gauge symmetries due to the singularity structure of the
manifold. 
For a particular chain of CY spaces ending in 
$\IP_4^{(1,1,n,2n+4,3n+6)}[6n+12]$ and henceforth labeled as type A,
subsequent work has confirmed the
expected behavior thereby lending strong support to the duality conjecture
\cite{asp, km, mv, cf, bkkm, bikmsv}.

Based on the pattern of weight regularities, besides the A class, 
different classes B, C, \dots\ of dual type II chains were postulated in 
\cite{afiq} even though the heterotic models were not found at the time. 
Here we will show explicitly how to construct the required B, C, \dots\ 
heterotic chains. We will see that these 
models have an intrinsic six-dimensional description
in terms of $K3$ compactifications with non-semisimple $E_8\times E_8$
backgrounds. In fact, A, B, C, \dots\ models can be built up by embedding 
none, one, two, \dots\ $U(1)$ backgrounds in each $E_8$ factor.
An interesting feature of the new models is the presence of anomalous
$U(1)$'s, that acquire mass at tree-level by swallowing zero modes of
the antisymmetric $B_{MN}$ field, together with non-anomalous
$U(1)$'s whose breaking corresponds to transitions $\cdots {\rm C} \to
{\rm B} \to {\rm A}$.

F-theory has  proved to be very fruitful for a geometric understanding of 
different string dualities \cite{vafa, mv, sen}.
In particular it was argued in \cite{mv} that F-theory compactifications 
to six dimensions on certain elliptically fibered 
CY 3-folds are dual to certain heterotic compactifications on $K3$. 
Upon further toroidal compactification on $T^2$, type II/heterotic duality 
is naturally recovered. By extending the analysis of \cite{mv} 
we will be able to construct explicit F-theory duals for the new
heterotic models. Moreover, each class of models will be shown to be
associated to fibrations of different elliptic fibers over the base  
$\IF _n$, thus establishing a correspondence between elliptic fiber
on the F-theory side and $U(1)$ factors on the heterotic side. 
More precisely, A, B, C, \dots\ models correspond to elliptic fibrations 
where the elliptic fiber is respectively, $\IP_2^{(1,2,3)}[6]$, 
$\IP_2^{(1,1,2)}[4]$, $\IP_2^{(1,1,1)}[3]$, \dots\ .
We will also argue that from the point of view of type IIA compactifications
the change of elliptic fiber appears to correspond to conifold transitions
as suggested in \cite{mv}.

This article is structured as follows. In Chapter 2 
we introduce some basic concepts and notation and review the properties 
of the chains of models proposed in \cite{afiq}. In Chapter 3 we explore 
heterotic $K3$ and $K3\times T^2$ compactifications with 
generic background embeddings in $E_8\times E_8$ 
containing both Abelian and non-Abelian factors. 
A detailed case by case analysis of different assignments of instanton 
numbers indicates perfect agreement of the resulting spectra with the Hodge 
numbers of the CY chains of various types. New model building possibilities 
using semisimple non-Abelian backgrounds are also discussed and 
intrinsically four-dimensional
heterotic chains involving enhancing of the toroidal $U(1)$'s are considered
to some extent. In Chapter 4, F-theory duals are constructed. Different 
fibrations are studied and a singularity analysis is performed to 
identify enhanced gauge symmetries. Conifold transitions among different chains 
are also discussed. Chapter 5 is devoted to the study of transitions to models 
without tensor multiplets. Conclusions, miscellaneous results and general 
outlook are presented in Chapter 6.
 
\section{Heterotic/Type II Duality and $D=6$  Heterotic Dynamics}
\bigskip

Type IIA compactifications are
characterized by the CY Hodge numbers $(b_{21},b_{11})$ 
where $b_{21}+1$ counts the number of hypermultiplets (including the 
dilaton) and $b_{11}$ counts the number of vector multiplets. 
The perturbative gauge group including the graviphoton is $U(1)^{b_{11}+1}$. 
On the other hand, $N=2$ heterotic strings in general have gauge symmetry 
group $G$ of ${\rm rank}(G)=\nv+1$ including the graviphoton. Here $\nv$ 
counts the number of vector multiplets including the dilaton. Giving vevs 
to adjoint scalars in vector multiplets realizes the transition to the 
Coulomb phase in which the gauge group is generically broken to 
$U(1)^{\nv+1}$. A necessary requirement for duality is therefore 
$(b_{21},b_{11})=(\nh-1,\nv)$, where $\nh$ is the number of 
(neutral) hypermultiplets that remain massless in the Coulomb phase.
It is also required that the candidate CY dual be a $K3$ fibration
\cite{klm, vw, al}.

In ref.~\cite{afiq} different models were constructed mainly by considering
$T^4/Z_M$ ($M=2,3,4,6$) orbifold limits of $K3$ and by embedding the orbifold 
action as a shift in the $E_8\times E_8$ or $Spin(32)/Z_2$ gauge lattice. 
After compactification on $T^2$, $N\!=\!2$, $D\!=\!4$ models were obtained.
The rank of the starting gauge group was then reduced in steps by giving 
vevs to scalar in hypermultiplets. Moving to the Coulomb phase at each step, 
$(\nh-1,\nv)$ was compared with candidate $CY$ Hodge numbers. 
This produced the chains of models of table 1 in \cite{afiq}. We will 
refer to these as chains of A type. 

A unified and extended version of these A models can be obtained by 
considering $K3$ compactifications with instanton backgrounds in 
$E_8\times E_8$. Let us denote by $H_{1,2}$ the background gauge (simple) 
groups and by $(k_1,k_2)$ the corresponding instanton numbers.
{}From each $E_8$ the unbroken gauge group is the commutant $G_i$ 
of $H_i$. The adjoint representation of $E_8$ decomposes 
under  $G\times H$ as ${\bf 248}$$=\sum_a (R_a,M_a)$.  
The number of  hypermultiplets in the representation $R_a$ of 
the unbroken group $G$ is then computed from the index theorem 
\beq
N(R_a) = k\ T(M_a)\ -\ {\rm dim}\, M_a
\label{nrabund}
\eeq
where  $T(M_a)$ is  given by $\tr \,(T_a^i T_a^j) = T(M_a) \delta _{ij}$,
$T_a^i$ being an $H$ generator in the representation $M_a$ 
\footnote{ Our normalization is such that $T(fund) = \oh, 1, 3,6$ and 
$30$ for $SU(N)$, $SO(N)$, $E_6$, $E_7$ and $E_8$ respectively.}. 
For example, an $SU(2)$ bundle with instanton number $k$ gives
$(k-4)/2$ multiplets in the ${\bf 56}$ of  $E_7$ and  $(2k-3)$ singlets

Since anomaly cancellation requires $k_1+k_2=24$,
it is convenient to define 
\beq
k_1=12+n \quad\quad ; \quad\quad     k_2= 12-n
\label{ndef}
\eeq
and, without loss of generality, assume $n\ge 0$. For $n \leq 8$
it is possible to have an $E_7\times E_7$ 
unbroken gauge group with hypermultiplet content 
\beq
\oh(8+n)({\bf 56},{\bf 1}) + \oh (8-n)({\bf 1}, {\bf 56})
+  62({\bf 1}, {\bf 1})
\label{e7e7}
\eeq
Due to the pseudoreal character of the ${\bf 56}$ of $E_7$,
odd values of $n$ can also be considered.
For $9 \leq n \leq 12$, $k_2$ is not large enough to support an  
$SU(2)$ background. 
The instantons become small and turn into
$12-n$ extra tensor multiplets. The unbroken gauge group is 
now $E_7 \times E_8$ with matter content 
\beq
\oh(8+n)({\bf 56},{\bf 1}) + (53+n)({\bf 1}, {\bf 1})
\label{e7e8}
\eeq
Models with various groups can be obtained from (\ref{e7e7})
and (\ref{e7e8}) by symmetry breaking.  

Notice that the group from the second $E_8$
does not posses, in general, enough charged matter to be completely broken.
Higgsing stops at some terminal group, depending on the value of $n$, 
with minimal or no charged matter. For instance $E_8$, $E_7$, $E_6$, $SO(8)$ 
terminal groups are obtained for $n=12,8,6,4$ while complete breaking 
proceeds for $n=2,0$. On the other hand, the first $E_7$ can be completely 
Higgsed away. The type A chains in \cite{afiq} are reproduced by cascade 
breaking through 
\beq 
\cdots \rightarrow SU(4) \rightarrow SU(3) \rightarrow SU(2) \rightarrow 
\emptyset 
\label{casca}
\eeq
In these chains, the weights of the candidate dual CY hypersurfaces 
in projective space follow a well defined pattern of regularities. Namely, 
the cascade Higgsing (\ref{casca}) maps into the following sequence 
in the type II side 
\beqa
& {} & \IP_5^{(1,1,w_1,w_2,w_3,w_4)}  \rightarrow  
\IP_4^{(1,1,w_1,w_2,w_3)} \rightarrow 
\nonumber \\
&{}& \hspace{1cm} 
\IP_4^{(1,1,w_1,w_2,w_3+w_1)} \rightarrow 
\IP_4^{(1,1,w_1,w_2+w_1,w_3+2w_1)} 
\label{cypatt}
\eeqa
where for the $\IP_4$'s the degree of the hypersurface 
defining equation is of course
the sum of the weights and for the $\IP_5$ there are two equations of
appropriate degrees. Moreover, these transitions can be recast in terms 
of $n$. In fact, in A models the last steps of cascade Higgsing
(\ref{casca}) have candidate CY duals
\beqa
& {} & \IP_5^{(1,1,n, n+4,n+6, n+8)}[2n+12,2n+8]  
\rightarrow  \IP_4^{(1,1,n,n+4,n+6)}[3n+12] \rightarrow 
\nonumber\\
&{}& \hspace{0.5cm}
\IP_4^{(1,1,n,n+4,2n+6)}[4n+12] \rightarrow 
\IP_4^{(1,1,n,2n+4,3n+6)}[6n+12] 
\label{achain}
\eeqa
This structure also holds for odd values of $n$ \cite{cf}.

Encouraged by these regularities, new chains were proposed in \cite{afiq}
by reorganizing $K3$ fibrations in the list of \cite{klm} according to eq. 
(\ref{cypatt}), even if heterotic models were not known at that time. 
For example, there exist chains of $K3$ fibrations, in which $b_{11}$
jumps by one in each step, that end at the
penultimate stage in (\ref{cypatt}) and that for generic $n$ have the form
\beqa
& {} &  \IP_5^{(1,1,n,n+2,n+4,n+6)}[2n+8,2n+6]  
\rightarrow \IP_4^{(1,1,n,n+2,n+4)}[3n+8] \rightarrow
\nonumber\\
&{}& \hspace{1.5cm}
\IP_4^{(1,1,n,n+2,2n+4)}[4n+8] 
\label{bchain} 
\eeqa
We will refer to these as chains of models B. There are also chains
of models C that have two elements given by
\beq
\IP_5^{(1,1,n,n+2,n+2, n+4)}[2n+6,2n+4]  \rightarrow
\IP_4^{(1,1,n,n+2,n+2)}[3n+6] 
\label{cchain} 
\eeq
Finally, there are chains of type D with single element
\beq 
\IP_5^{(1,1,n,n+2,n+2, n+2)}[2n+4,2n+4]
\label{dchain}
\eeq
The structure of the CY chains is summarized in Table~\ref{tabla1}.
In each case $n$ is restricted by the condition that the set of weights
lead to a well defined CY space. For type A, $n \leq 12$ in agreement
with the heterotic construction. For types B and C, the weights
correspond to reflexive polyhedra only for $n \leq 8$ and $n \leq 6$ 
respectively. For models D, $n \leq 4$ is expected.

\begin{table}[htb]
\renewcommand{\arraystretch}{1.25}
\begin{center}
\begin{tabular}{|c|l|l|}
\hline
$r$ & \hspace{1.4cm}A \quad \quad $2 \leq n \leq 12$ &  
\hspace{1.4cm}B \quad\quad $2 \leq n \leq 8$  
\\[0.2ex] \hline
4 & $\IP_5^{(1,1,n, n+4,n+6, n+8)}[2n+12,2n+8]$ & {} \\[0.2ex]
3 & $\IP_4^{(1,1,n,n+4,n+6)}[3n+12]$ &
$\IP_5^{(1,1,n,n+2,n+4,n+6)}[2n+8,2n+6]$ \\[0.2ex]
2 & $\IP_4^{(1,1,n,n+4,2n+6)}[4n+12]$ &
$\IP_4^{(1,1,n,n+2,n+4)}[3n+8]$ \\[0.2ex]
1 & $\IP_4^{(1,1,n,2n+4,3n+6)}[6n+12]$ & 
$\IP_4^{(1,1,n,n+2,2n+4)}[4n+8]$  \\[0.2ex]
\hline
\hline
$r$ &\hspace{1.4cm}  C \quad\quad  $2 \leq n \leq 6$ & 
\hspace{1.4cm} D \quad\quad $2 \leq n \leq 4$  
\\[0.2ex] \hline
2 & $\IP_5^{(1,1,n,n+2,n+2, n+4)}[2n+6,2n+4]$ & {} \\[0.2ex]
1 & $\IP_4^{(1,1,n,n+2,n+2)}[3n+6]$ & 
$\IP_5^{(1,1,n,n+2,n+2, n+2)}[2n+4,2n+4]$ \\[0.2ex] 
\hline
\end{tabular}
\end{center}
\caption{ Structure of the A,B,C and D chains.}
\label{tabla1}
\end{table}

The Hodge numbers for the terminal elements of each chain are given
in Table~\ref{tabla2} for future reference
\footnote{Most of these results, as well as those in eq.~(\ref{abchodge}) 
below,  appear in refs.~\cite{ks, klm}. The remaining 
cases in $\IP_4$ 
have been computed using the program {\tt POLYHEDRON}
written by P.~Candelas. The numbers for the spaces in $\IP_5$ 
were calculated by A.~Klemm.}.
The expressions in Table~\ref{tabla1} clearly do not apply to $n=0$ nor
to $n=1$, since it is known, for instance, that $\IP_4^{(1,1,1,6,9)}[18]$
is not a $K3$ fibration. However, these two values are naturally 
considered once we notice that the terminal spaces correspond to
elliptic fibrations over $\IF_n$ that can be extended to $n=0,1$
using the formalism of ref.~\cite{mv}. In section 4 we will
explain in more detail the elliptic fibration structure of
the various models. It is also worth noticing that the Hodge numbers
$(b_{12}^r, b_{11}^r)$ for the chains in Table~\ref{tabla1} can all
be written in terms of the $(b_{12}^1, b_{11}^1)$ recorded in
Table~\ref{tabla2}. Specifically,
\beqa
{\rm A} : \quad \quad 
              b_{12}^2 & = & b_{12}^1 - (12n+29) \nonumber \\[0.2ex]
              b_{12}^3 & = & b_{12}^1 - (18n+46) \nonumber \\[0.2ex]   
              b_{12}^4 & = & b_{12}^1 - (22n+61) \nonumber \\[0.4ex] 
{\rm B} : \quad\quad 
              b_{12}^2 & = & b_{12}^1 - (6n+15)  \nonumber \\[0.2ex]
              b_{12}^3 & = & b_{12}^1 - (10n+26) \nonumber \\[0.4ex]
{\rm C} : \quad\quad 
              b_{12}^2 & = & b_{12}^1 - (4n+11) 
\label{abchodge}
\eeqa
In all cases $b_{11}^r = b_{11}^1 + r-1$.
These results are tabulated in Table~\ref{tabla3}, 
at the end of the article, for the reader's
convenience. 

\begin{table}[htb]
\small
\renewcommand{\arraystretch}{1.25}
\begin{center}
\begin{tabular}{|c|cccc|}
\hline
$n$ & A & B & C & D\\[0.2ex]
\hline
\hline
0 & (243,3) & (148,4) & (101,5) & (70,6) \\
1 & (243,3) & (148,4) & (101,5) & (70,6) \\
2 & (243,3) & (148,4) & (101,5) & (70,6) \\
3 & (251,5) & (152,6) & (103,7) & (70,10) \\
4 & (271,7) & (164,8) & (111,9) & (76,10) \\
5 & (295,7) & (178,10) & (120,12) & {} \\
6 & (321,9) & (194,10) & (131,11) & {} \\
7 & (348,10) & (210,12) & {} & {} \\
8 & (376,10) & (227,11) & {} & {} \\
9 & (404,14) & {} & {} & {} \\
10 & (433,13) & {} & {} & {} \\
11 & (462,12) & {} & {} & {} \\
12 & (491,11) & {} & {} & {} \\
\hline 
\end{tabular}
\end{center}
\caption{ Hodge numbers $(b_{21}^1, b_{11}^1)$ for the terminal spaces .}
\label{tabla2}
\end{table}
  
In section 3 we will develop the heterotic construction that
reproduces systematically the terminal elements of type B, C and D.
Moreover, we will show how un-Higgsing of $SU(r)$ factors
leads to spectra that match the Hodge numbers of the chains
given in (\ref{abchodge}). In the heterotic construction many more 
symmetry breaking patterns are possible. We then expect that the terminal 
CY spaces are continuously connected to points with generic enhanced
gauge symmetries as shown recently for the A models \cite{cf, bikmsv}. 

It is not clear from the preceding discussion if there exists
any correspondence among models with same value of $n$. 
However, the results in Table~\ref{tabla2} suggest
that this is indeed the case. For instance, the 
$n=4$, $(271,7)$ model in chain A with $SO(8)$ terminal group 
corresponds to the $n=4$ models $(164,8)$ in chain B, $(111,9)$ in 
chain C and $(76,10)$ in chain D. We observe that the rank increases in one 
unit when  A $\to$ B $\to$ C $\to$ D. On the other hand, the number 
$\nh$ of hypermultiplets decreases in each step. 
This can be taken as an indication of the presence of an extra 
$U(1)$ group for models B so that their unbroken gauge group would be 
$SO(8)\times U(1)$. Likewise, there would be two and three extra $U(1)$ 
factors for models C and D. The existence of charged matter with respect 
to these $U(1)$ groups would explain the decreasing in $\nh$. Similar arguments apply to other $n$'s. 
For values such as $n=5$ our heterotic construction will also explain the horizontal behavior of $b_{11}$.

In the type A heterotic models the gauge 
group structure before going to the Coulomb branch is of the form 
$G\times U(1)^4$ where $U(1)^4$ arises in the toroidal compactification from 
six to four dimensions. Since $T^2$ is untouched we can interpret 
the A models as intrinsically corresponding to $N\!=\!1$ compactifications 
on $K3$. We will see that this is also the case for models B,C and D. 
It is then useful to recall some properties of $N=1$ six-dimensional 
theories that are in a sense more constrained since being 
chiral they could have potential anomalies. In particular, the anomaly 
8-form should factorize as 
\beq
I_8=
(\tr\, R^2 -v_{\alpha }\tr\, F_\alpha^2 )
(\tr \, R^2 -{\tilde v}_{\alpha }\tr\, F_\alpha^2 )
\label{anfact}
\eeq
where $\alpha$ runs over the gauge factors.
The coefficients $v_{\alpha }$ are fixed for each gauge group. They are given 
by  $v_{\alpha }=2,1,\frac13,\frac13,\frac16,\frac{1}{30}$ for  
$SU(N), SO(N), F_4, E_6,E_7, E_8$ \cite{erler} for Kac-Moody level one.
On the other hand, the ${\tilde v}_{\alpha }$ coefficients 
depend on the hypermultiplet content of each group. 
For instance ${\tilde v }_{E_7 } = \frac16(n_{56}-4)$. Results for other 
groups can be found in ref.~\cite{erler}. For generic
gauge group $G= G_1\times G_2$ with $G_1$ and $G_2$ subgroups of the first and 
second $E_8$ obtained from backgrounds with instanton numbers $(12+n,12-n)$,
the following identity is satisfied 
\beq
\frac{{\tilde v}_{1}}{v_1}= \frac{n}{2}  \quad\quad ; \quad\quad   
\frac{{\tilde v }_{2}}{v_2}= -\frac{n}{2} 
\label{vtfor}
\eeq
These relations remain valid at each step of possible Higgsing.

{}From the anomaly polynomial it follows that the gauge kinetic 
terms are proportional to \cite{sagnotti}
\beq
-v_1 (e^{-\phi}+ \frac{n}2e^{\phi})\tr\, F_1^2 -
 v_2 (e^{-\phi}-\frac{n}2e^{\phi})\tr\, F_2^2 
\label{sagno}
\eeq
where $F_i$ is the field strength of the unbroken groups $G_i$ 
and $\phi$ is the scalar dilaton living in a 6d tensor multiplet.
Heterotic/heterotic duality  \cite{dl} is obtained
for $n=0$ if small instanton effects are taken into
account \cite{dmw} . It is also present in the $n=2$ case
\cite{afiq2}  if one Higgses  away the second group factor. 
In fact both cases $n=0$ and $n=2$ turn out to be 
connected if examined from the F theory point of view
\cite{mv}. 
The coefficient of the gauge kinetic term for the second $E_8$ is
such that the gauge coupling diverges at \cite{dmw, sw} 
\beq
e^{-2\phi} = \frac{n}2
\label{phtr}
\eeq
This is a sign of a phase transition in which there appear tensionless
strings \cite{gh, sw, dlp}  .  

In the previous discussion of six-dimensional heterotic strings we have 
generically assumed the presence of just the dilaton tensor multiplet. 
This is in fact the correct description at a perturbative level.
However, in general six-dimensional $N=1$ theories more than 
one tensor multiplet may be present. Indeed, compactifications of 
$M$-theory on $K_3\times S^1/Z_2$ leads to this possibility not 
seen at the perturbative level \cite{dmw, sw, witfiveb}. 
In fact, five-branes located at points (parametrized by five real 
coordinates) in this internal
space will be generically present. A tensor and a hypermultiplet  
are associated  to these branes. The five-branes are a source of torsion
so that in a case with $k_1$ instantons in the first $E_8$, $k_2$ 
in the second and $n_T-1$ five-branes at points in 
$K_3\times S^1/ Z_2$, the condition $k_1 + k_2 = 24$ is replaced by 
\beq
k_1+k_2+n_T-1=24
\label{ancant}
\eeq
Here $n_T$ is the number of tensor multiplets including that of the
dilaton. Moreover, cancellation of gravitational anomalies leads to  
\beq
n_H + 29 n_T- n_V = 273
\label{gravan}
\eeq
This equation is for example satisfied by (\ref{e7e8}) since
$\nt-1=12-n$.

Before getting into the specific discussion of 6d models either from the 
heterotic side or from an F-theory approach, let us recall that there are 
still heterotic/type II dual candidates
that are not understandable from a six-dimensional point of view.  
The heterotic version of these models would require, in general, introduction
of asymmetric orbifolds or enhancings involving the two-torus appearing 
when compactifying   to $D=4$ (or from $D=10$ to $D=8$, followed 
by a $K3$ compactification). This is the case 
for instance for the $(128,2) \equiv \IP_4^{(1,1,2,2,6)}[12]$ 
model discussed in \cite {kv}. Such models  may 
still be organized into chains according to eq.~(\ref{cypatt}),  
as was noticed in \cite{afiq}. For example,
$(76,4) \rightarrow (99,3) \rightarrow (128,2)$. 
Also $(75,9) \rightarrow (104,8)  \rightarrow (143,7)  \equiv 
\IP_4^{(1,1,4,4,10)}[20]$ 
follows the same pattern. These cases do not correspond to
elliptic fibrations. A possible scheme for obtaining the  heterotic  
candidates some of these  4d chains is discussed at the end of next 
chapter.

\section{Heterotic Strings on $K3$ and $K3 \times T^2$ Revisited} 
\label{sec:hetk3}
\bigskip
\subsection{Non-semisimple Backgrounds}
\label{sub:u1s}

As we discussed in chapter 2, finding the heterotic duals
of the chains of Calabi-Yau models of type B,C, \dots \ seems to require
new ingredients beyond the usual instanton embedding in $E_8\times E_8$.
In this chapter we will show how the desired new 6d heterotic models 
can be most readily obtained by considering generic 
$H\times U(1)^{8-d}$  backgrounds in each $E_8$, with $H$ some 
non-Abelian factor.

$U(1)$ backgrounds on $K3$ were first explored by Green, Schwarz and 
West \cite{gsw}. The procedure can be applied to $SO(32)$ or
$E_8\times E_8$ heterotic strings. We will concentrate in the latter
case and to begin we consider $U(1) \subset E_8$.
The instanton number of the $U(1)$ configuration is defined to be
\beq
m_i\ =\ \frac{1}{16\pi^2}  \int_{K3} \frac{1}{30} \Tr\, F^2_{U(1)_i} \quad ; 
\quad i=1,2 
\label{mdef}
\eeq
Considering both $E_8$'s and imposing the requirement 
$\int_{K3} (\tr \, R^2 - \frac1{30}\Tr \, F^2) =0$ gives the condition
$m_1 + m_2 = 24$. 

To determine the matter spectrum it is convenient
to consider separately each $E_8$ broken to $E_7 \times U(1)$. The
relevant adjoint decomposition is
\beq
{\bf 248} = ({\bf 133}, 0) + ({\bf 56}, q) + ({\bf 56}, -q) +
({\bf 1}, 2q) + ({\bf 1}, -2q) + ({\bf 1}, 0)
\label{e7u1}
\eeq
Here $q =\oh$ so that the $U(1)$ generator $Q$ is normalized as a generator 
of $E_8$ in the adjoint representation,
{\it i.e.} $\Tr \, Q^2 = 30$. According to the index theorem
(\ref{nrabund}), the number of hypermultiplets of charge $q$ is simply 
$N_q = mq^2 - 1$. Then, 
\beqa
N({\bf 56}, \oh) & = & N({\bf 56}, -\oh)= \frac{m}4 - 1 \nonumber\\
N({\bf 1}, -1) & = &  N({\bf 1}, -1) = m \ -1 
\label{partun}
\eeqa
Notice that to obtain positive multiplicities and half-integer number of 
${\bf 56}$'s, $m$ must be an even integer with $m\geq 4$. 
Taking into account both $E_8$'s, the allowed values for $(m_1,m_2)$ 
are $(24,0)$, $(20,4)$, $(18,6)$, $(16,8)$, $(14,10)$  and $(12,12)$. The 
hypermultiplet content
of these $E_7 \times U(1) \times E_7 \times U(1)$ models is
\beqa
&{}& \big \{
\frac14 (m_1 - 4) ({\bf 56}, \oh ; {\bf 1},0)   + 
\frac14 (m_2 - 4) ({\bf 1}, 0 ; {\bf 56},\oh) + 
\nonumber \\
&{}& (m_1 - 1) ({\bf 1}, 1 ; {\bf 1},0)  +   
(m_2 - 1) ({\bf 1}, 0 ; {\bf 1}, 1) + {\rm c.c.} \big \} + 
20 ({\bf 1}, 0 ; {\bf 1},0) 
\label{hyperuno}
\eeqa
where we have added the gravitational contribution.

It is easy to check that $U(1)$'s in this class of theories are in 
general anomalous. More precisely, one finds that the anomaly 8-form 
$I_8$  does not generically factorize into a product of two 4-forms 
so that the Green-Schwarz mechanism  cannot  cancel  the residual anomaly.
Instead one finds that the linear combination of $U(1)$ charges
\beq
 Q_f  = \cos\theta \, Q_1  + \sin\theta \, Q_2   
\label{lacomb}
\eeq
leads to a factorized $I_8$  as long as
\beq
\sin^2\theta  = {{m_2}\over {m_1+m_2}} \quad\quad ; \quad\quad    
\cos^2\theta  = {{m_1}\over {m_1+m_2}} 
\label{normal}
\eeq
Thus, for given $m_{1,2}$, there is a linear combination of both $U(1)$'s
which is anomaly-free  but the orthogonal combination is not.
Thus, somehow, the latter combination must be spontaneously broken.  
Indeed, a mechanism by which this can take place was suggested in 
refs.~\cite{witso, gsw} for analogous compactifications. The idea is 
that in $D=10$ the kinetic term of the $B_{MN}$ field contains a piece
\beq
H^2 \simeq  (\, \partial _{\mu }B_{ij}\ +\ A^1_{\mu }
\langle F^1_{ij}\rangle \ + \  \ A^2_{\mu }\langle F^2_{ij}\rangle \, )^2 
\label{truco}
\eeq
where the indices  $i,j$ live in the four compact dimensions. Notice that one  
linear combination of $A^1_{\mu }$ and $A^2_{\mu }$  will become massive 
by swallowing a $B_{ij}$ zero mode. 
Specifically, since $\langle F^a_{ij}\rangle \simeq \sqrt{m_a}$ 
($a=1,2$), precisely the orthogonal combination to that in
eq.(\ref{lacomb}) acquires a mass through this mechanism. Thus, 
the would-be anomalous $U(1)$ is in fact absent from the massless spectrum 
and the gauge group is actually  $E_7\times E_7\times U(1)_f$.
Notice that if we further break down $U(1)_f$ by giving vevs to some of the 
singlets in (\ref{hyperuno}), the hypermultiplet content of this 
class of models with $U(1)$ instanton numbers  $(m_1,m_2)$  is analogous 
to that obtained with $SU(2)$ instantons $(k_1,k_2)$ and
$k_i$ even. However, if the Higgs breaking proceeds through charged 
multiplets, at each stage of symmetry breaking there survives an unbroken 
$U(1)$ corresponding to a linear combination of $Q_f$ and $E_7$ Cartan 
generators. We will see below that this residual $U(1)$ has an important 
role in understanding the extra families of models baptized B, C and D in 
the previous section.

Clearly, it is also possible to combine Abelian and non-Abelian backgrounds 
and have, for example, $H\times U(1)$ bundles with instanton numbers $(k,m)$ 
inside each $E_8$. The conmutant is $G\times U(1)$ and the adjoint of 
$E_8$ decomposes as ${\bf 248}=\sum_a(R_a,q_a,M_a)$
under $G\times U(1)\times H \subset E_8$. 
The number of hypermultiplets in the representation $(R_a,q_a)$ of
$G\times U(1)$ is again given by the index theorem
\beq
N(R_a,q) =  kT(M_a) + mq^2 {\rm dim}\, M_a  - {\rm dim}\, M_a 
\label{hqbund}
\eeq
where we again normalize $\Tr \, Q^2=30$. The generalization 
to $H\times U(1)^{8-d}$,
$d \leq 6$, is straightforward. In the following we will consider various 
choices leading to the type B, C and D heterotic duals.
\subsubsection{Type B Models}
\label{sub:tipob}

Choosing $SU(2)\times U(1)$ as background gives an unbroken subgroup   
$E_6\times U(1)$ arising from each $E_8$. The relevant adjoint decomposition is
\beqa 
{\bf 248} & = & ({\bf 78},0,{\bf 1}) + ({\bf 1}, 0, {\bf 1}) + 
({\bf 1},0, {\bf 3}) + ({\bf 27},q,{\bf 2}) + (\ov{{\bf 27}},-q,{\bf 2}) +
\nonumber \\
& {} & (\ov{{\bf 27}}, 2q,{\bf 1}) + ({\bf 27},-2q,{\bf 1}) +
({\bf 1},3q,{\bf 2}) + ({\bf 1},-3q,{\bf 2}) 
\label{e6su2u1}
\eeqa
where now $q=\frac{1}{2\sqrt3}$. Embedding $SU(2)\times U(1)$ backgrounds  
with instanton numbers $(k_1,m_1;k_2,m_2)$ in both $E_8$'s gives the 
following $E_6\times U(1) \times E_6\times U(1)$ spectrum
\beqa
& {} &\big \{
\ds{\frac16(3k_1 + m_1 -12)} \ds{({\bf 27},\frac1{2\sqrt3};{\bf 1},0) +}  
\ds{\frac16(3k_2 + m_2 -12)} \ds{({\bf 1},0 ; {\bf 27},\frac1{2\sqrt3})+}
\nonumber \\[2mm]
& {} & 
\ds{\frac13(m_1 -3)} \ds{({\bf 27},-\frac1{\sqrt3};{\bf 1},0) +}
\ds{\frac13(m_2 -3)} \ds{ ({\bf 1},0 ; {\bf 27},-\frac1{\sqrt3}) + } 
\nonumber\\[2mm]
& {} & 
\ds{\frac12(k_1 + 3m_1 -4)}\ds{({\bf 1},\frac{\sqrt3}{2};{\bf 1},0)} 
+ \ds{\frac12(k_2 + 3m_2 -4)} \ds{({\bf 1},0 ; {\bf 1},\frac{\sqrt3}{2}) +} 
{\rm c.c.} \big \} + 
\nonumber\\[2mm]
& {} &
\ds { ( (2k_1-3)+ (2k_2-3)}  
\ds { + 20 )({\bf 1},0 ; {\bf 1},0)} 
\label{espuna}
\eeqa
In this case gravitational anomalies cancel as long as 
$k_1+m_1+k_2+m_2=24$.  
Again, we find that, independently of the values of $k_{1,2}$,
the linear combination in eq.(\ref{lacomb}) leads to a factorized
$U(1)$ anomaly that can be cancelled by the GS mechanism.
The orthogonal linear combination is expected to be Higgsed away as
in the previous case.

Notice that  in the presence of the $SU(2)$ bundles the values of $m_{1,2}$  
are forced to be multiples of $3$ in order to have half-integer numbers of 
$({\bf 27} + \ov{{\bf 27}})$ and also $m_{1,2}\geq 3$.  
Thus the simplest class of models of this type will have instanton numbers 
$(k_1,3;k_2,3)$  and the unbroken $U(1)_f$ is in this case the diagonal 
combination $U(1)_D$. The $E_6\times E_6\times U(1)_D$ spectrum  is 
then given by
\beqa
& {} & \big \{
\frac12 (k_1 - 3) ({\bf 27},{\bf 1},\frac1{2\sqrt6}) +
\frac12 (k_2 -3) ({\bf 1},{\bf 27},\frac1{2\sqrt6}) +
\nonumber \\
& {} & \frac12 (k_1 + k_2 + 10) ({\bf 1},{\bf 1},\frac{\sqrt3}{2\sqrt2}) +
{\rm c.c.} \big \} + (2k_1 + 2k_2 + 13) ({\bf 1}, {\bf 1},0)
\label{espudos}
\eeqa
This matter content has anomaly polynomial
\beq
I_8=
(\tr\, R^2 - \frac 13 \tr\, F_1^2 - \frac 13 \tr\, F_2^2 - f^2 )
(\tr \, R^2 - \frac{k_1-9}6 \tr\, F_1^2 - \frac{k_2-9}6 \tr\, F_2^2 
- 3 f^2 )
\label{ane6e6}
\eeq
where $F_i$ is the field strength of the $i$-th $E_6$ and $f$ is that
of $U(1)_D$. Notice that the mixed $\tr F_1^2 \, \tr F_2^2$ term vanishes
by virtue of the constraint $k_1 + k_2 = 18$.

The fact that $k_1+k_2=18$, instead of $k_1+k_2=24$ in the case without $U(1)$
backgrounds, hints at the required heterotic duals  
of models of type B. Indeed, in these models, the
range for the values of $n$ is smaller ($n\leq 8$) and this is
probably the case here since the range for $k_{1,2}$ is also smaller.
Moreover, models B have a number of vector 
multiplets one unit higher compared to the corresponding chain A 
elements. This is precisely the case here, due to
the presence of the extra $U(1)_D$. 
These arguments are compelling enough to consider this sort of
heterotic constructions in more detail. We will see that
upon sequential Higgsing of the non-Abelian symmetries the spectrum
in (\ref{espudos}) does in fact reproduce chains of type B. 

In analogy with the usual situation, we will label the 
models in terms of the integer
\beq
n  =  k_1\ +\ m_1 \ - \ 12
\label{nuevon}
\eeq
where we assume without loss of generality that  $k_1+m_1\geq 12$.  We
choose $m_1=m_2=3$ as before so that $k_1+k_2= 18$ (we will show that  
for $n=7$  the $k_2$ instantons become small).
We now set up the derivation of the spectrum implied by (\ref{espudos})
upon maximal Higgsing of non-Abelian symmetries. The results of course
depend on $n$ or equivalently on the pair $(k_1,k_2)$. The strategy is
to first implement breaking of the second $E_6$ together with $U(1)_D$
to $G_0 \times U(1)_X$, where $U(1)_X$ is 
the appropriate `skew' combination of $U(1)_D$
and an $E_6$ Cartan generator. Since $k_1 \geq 9$, the first $E_6$ together
with $U(1)_X$ can then be broken to another `skew' $U(1)_Y$. The terminal
gauge group is therefore $G_0 \times U(1)_Y$ which by construction has
a factorized anomaly polynomial. Except for $n=5$, the terminal matter
consists of $G_0$ singlets charged under $U(1)_Y$ plus a number of
completely neutral hypermultiplets. The final step is to perform a
toroidal compactification on $T^2$ followed by transition to the 
Coulomb phase. This allows us to compare the resulting spectrum of
vector and hypermultiplets with the Hodge numbers of candidate dual
type II compactifications. It is also interesting to consider
un-Higgsings in the first $E_6$ along different branches. We will 
study in particular un-Higgsing of $SU(r)$ factors in order to identify
the type B chains more precisely.
We now sketch the outcome for the different allowed values of  $n$.

\bigskip\noindent 
$n=0,1,2$   
\medskip

\noindent
In these cases $(k_1,k_2)=(9,9), (10,8)$ and $(11,7)$ respectively
and the terminal group is just $U(1)_Y$. The terminal hypermultiplet
content is
\beq 
\{ 48(\frac1{2\sqrt2}) + {\rm c.c.} \}  
+ 149(0)
\label{ncud}  
\eeq
in all three cases. In the 4d Coulomb phase there are then 4 vector 
multiplets and 149 massless hypermultiplets. This implies Hodge numbers 
$(b_{21},b_{11})=(148,4)$ in agreement with the values for B models
given in Table~\ref{tabla2}.
{} From the spectrum (\ref{ncud}) it is obvious that if we further
break $U(1)_Y$ we end up with 244 hypermultiplets and no 6d vector
multiplets, corresponding to the final elements of the $n=0,1,2$ A chains.
Although these three cases yield similar spectra after full Higgsing of
$E_6 \times E_6$, if we un-Higgs in steps it is easy to see that they
behave differently. For example, un-Higgsing an $SU(2)$ factor in the
first $E_6$ gives $SU(2)\times U(1)_Y$ spectrum
\beq 
\{ (3n+8)({\bf 2}, \frac1{2\sqrt7}) + 
(41-3n)({\bf 1}, \frac1{\sqrt7})+ {\rm c.c.} \}  
+ (134-6n)({\bf 1}, 0)
\label{ncudsu2}  
\eeq
The number of vector and hypermultiplets in the 4d Coulomb phase clearly
matches the Hodge numbers given in (\ref{abchodge}).

\bigskip\noindent
$n=3$  
\medskip

\noindent
Here $(k_1,k_2)=(12,6)$ and the terminal group is
$SU(3)\times U(1)_Y$ with hypermultiplets transforming as
\beq
\{ 50 ({\bf 1}, \frac{1}{2} \sqrt{\frac{3}{5}}) + {\rm c.c.} \} + 
153({\bf 1}, 0)
\label{ntres} 
\eeq
Upon toroidal compactification and transition to the 4d Coulomb phase
the spectrum matches the Hodge numbers $(b_{21},b_{11})=(152,6)$
characteristic of the space $\IP_4^{(1,1,3,5,10)}[20]$. Notice again
that Higgsing $U(1)_Y$ leads to the final element of the $n=3$ type A chain 
with $(b_{21},b_{11})=(251,5)$.

\bigskip\noindent
$n=4$
\medskip

\noindent
In this case $(k_1,k_2)=(13,5)$ and the terminal group is
$SO(8)\times U(1)_Y$ with hypermultiplet content
\beq 
\{ 54({\bf 1}, \frac1{\sqrt{6}}) + {\rm c.c.} \}  
+ 165({\bf 1}, 0)
\label{ncuatro}  
\eeq
In the 4d Coulomb phase we obtain a spectrum with type IIA dual 
characterized by $(b_{21},b_{11})=(164,8)$. These are the Hodge numbers of
$\IP_4^{(1,1,4,6,12)}[24]$. Observe the presence of an enhanced $SO(8)$ 
group just as it happens in the $n=4$ type A terminal model.

\bigskip\noindent
$n=5$
\medskip

\noindent
In this case we have $(k_1,k_2)=(14,4)$ und upon Higgsing we arrive at a 
gauge group $E_6\times U(1)_Y$ with hypermultiplets transforming as
\beq
\{ {1\over 2}({\bf  {27}},\frac1{4\sqrt{3}} )+ 
{{117}\over 2}({\bf 1}, \frac{\sqrt{3}}{4}  )+ {\rm c.c.} \} \ + \
 179({\bf 1},0) 
\label{ncinco}
\eeq
Due to the $U(1)$ charge, the gauge symmetry cannot be 
further broken (as long as the $U(1)$ remains unbroken). 
Going to the 4d Coulomb phase we arrive at a model  
corresponding to Hodge numbers $(b_{21},b_{11})=(178,10)$.

\bigskip\noindent
$n=6$
\medskip

\noindent
Here $(k_1,k_2)=(15,3)$ so that the second $E_6$ has no charged matter
and cannot be broken. The terminal group is then $E_6\times U(1)_Y$ 
with hypermultiplet content
\beq 
\{ 64({\bf 1}, \frac{\sqrt{3}}{4}  ) + {\rm c.c.} \}  
+ 195({\bf 1}, 0)
\label{nseis}  
\eeq
Recall that the $n=6$ type A chain has a matter-free $E_6$ as terminal group. 
In the case at hand, going to the 4d Coulomb phase implies a dual with 
$(b_{21},b_{11})=(194,10)$ in agreement with the Hodge numbers of
$\IP_4^{(1,1,6,8,16)}[32]$.

\bigskip\noindent
$n=7$
\medskip

\noindent
Naively we would set $k_2=2$ but this does not lead to sensible results
as it is obvious from eq.(\ref{espudos}). We will then remove the $k_2$
$SU(2)$ instantons so that we are left with just an $U(1)$ background
in the second $E_8$. {} From eq.~(\ref{partun})
we see that we need to have $m_2\geq 4$, $m_2$ even. Hence, one of the 
two instantons in $k_2$ is employed in raising $m_2$ from 3 to 4  
and the other becomes point-like.  We then have a distribution
$(k_1,m_1;k_2,m_2)=(16,3;0,4)$, plus a point-like instanton giving rise to
a tensor multiplet.  Cancellation of gravitational anomalies is guaranteed
since $k_1+m_1+m_2+1=24$, where the extra 1 is due to the presence of an 
M-theory 5-brane (which in term gives rise to a 6d tensor multiplet).
The $E_7$ arising from the second $E_8$ has no charged matter as implied
by (\ref{partun}). Then, the terminal group is
$E_7\times U(1)_Y$ with hypermultiplets transforming as
\beq
\{ 69({\bf 1}, \frac{1}{\sqrt{5}}) + {\rm c.c.} \}  
+ 210({\bf 1}, 0)
\label{nsiete}
\eeq
To these we must add one neutral hypermultiplet and one 
tensor multiplet whose 5 scalar components parametrize the position
of the 5-brane in $K3\times S^1/Z_2$.
Upon further toroidal compactification the tensor multiplet gives rise to a
4d vector multiplet.  Moving to the 4d Coulomb branch we land on a
model that matches $(b_{21},b_{11})=(210,12)$. These are the Hodge numbers
of $\IP_4^{(1,1,7,9,18)}[36]$.

\bigskip\noindent
$n=8$
\medskip

\noindent
We now have $(k_1,m_1;k_2,m_2)=(17,3;0,4)$ and, unlike the previous
situation, there is no small instanton. The gauge group upon
Higgsing is $E_7\times U(1)_Y$ with hypermultiplets transforming as
\beq 
\{ 75({\bf 1}, \frac{1}{\sqrt{5}}  ) + {\rm c.c.} \}  
+ 228({\bf 1}, 0)
\label{nocho}  
\eeq
In the 4d Coulomb branch we find a model dual to a type IIA compactification 
on a CY with  $(b_{21},b_{11})=(227,11)$. These are precisely the Hodge
numbers of $\IP_4^{(1,1,8,10,20)}[40]$, the last element of 
the $n=8$ type B chain.

It is now easy to understand from the heterotic side why  $n\leq 8$.  
The next element in the series would have $k_1=18$  and $m_1=3$.  For the 
second $E_8$ the only sensible alternative consistent with anomaly 
cancellation is to have 3 point-like instantons.
The terminal gauge group is then  $E_8\times U(1)_Y$ with hypermultiplets
transforming as
\beq
\{ 78({\bf 1},\oh ) + {\rm c.c.} \}  
+ 246({\bf 1}, 0)
\eeq
There are in addition 3 tensor multiplets and 3 hypermultiplets whose scalar 
components parametrize the positions of the three 5-branes. In this
situation the $U(1)$ is actually anomalous and it is thus Higgsed away. 
Therefore, we are left altogether with 405 hypermultiplets.
In four dimensions  the enhanced gauge symmetry is
$E_8\times U(1)^3_{tensor} \times U(1)^4$, implying a Coulomb branch 
with 14 vector multiplets and 405 hypermultiplets. The dual then
corresponds to a type IIA compactification on a CY 
with Hodge numbers (404,14).  This is nothing but the final 
element of the $n=9$ type A chain. Hence, type B chains 
stop at $n=8$ because for $n\geq 9$ they fall back into type A chains.

The foregoing heterotic construction not only matches the Hodge numbers 
of the last elements of chains of type B but also reproduces the preceding 
elements in each of the chains. Indeed, considering the  
symmetry breaking sequence $SU(3)\times U(1)\rightarrow SU(2)\times 
U(1)$$\rightarrow U(1)$, and going to the 4d Coulomb phase, we find 
results in complete agreement with (\ref{abchodge}) for all $n$.
We can also consider other un-Higgsing patterns. For example, for
$SU(4)\times U(1)$ the corresponding Hodge numbers are given by
\beq
b_{12}^4  =  b_{12}^1 - (12n+33) \quad\quad ; \quad\quad
b_{11}^4  =  b_{11}^1 + 3
\label{bsu4}
\eeq
Notice also that chains of type A and B  
are connected in the heterotic side by Higgsing of the $U(1)$  
present in the latter. Thus, at each step of chain B   
there is a Higgs branch connecting it to the corresponding step 
in the type A chain with same $n$.

\subsubsection{Type C and D Models}
\label{sub:tipoc}

We now consider $SU(2)\times U(1)^2$ backgrounds in each $E_8$. The $U(1)$'s 
are embedded according to the branchings $E_8\supset SO(10)\times SU(4)$  and
$SU(4)\supset SU(2)\times SU(2)_A \times U(1)_B$ 
$\supset SU(2)\times U(1)_A\times U(1)_B$.  
The distribution of instanton numbers is chosen to be
$(k_1,m_{1A},m_{1B};k_2,m_{2A},m_{2B})=(k_1,3,2;k_2,3,2)$, which can be shown 
to guarantee a consistent spectrum. Notice
that anomaly cancellation requires $k_1+k_2=14$  (in the absence of extra 
tensor multiplets from small instantons). The unbroken gauge group at 
the starting level is $SO(10)\times U(1)^2\times SO(10)\times U(1)^2$. 
In this case the diagonal combinations $Q_{AD} = \s2 (Q_{1A} + Q_{2A})$
and $Q_{BD} = \s2 (Q_{1B} + Q_{2B})$ are anomaly-free
whereas their orthogonal combinations are anomalous and are expected
to be Higgsed away by a mechanism analogous to that explained before.
The $SO(10)\times SO(10)\times U(1)_{AD}\times U(1)_{BD}$ 
hypermultiplets in the massless spectrum are  
\beqa
& {} & \big \{
\frac12 (k_1 - 3) ({\bf 16},{\bf 1},0, -\frac{1}4) +
\frac12 (k_2 -3) ({\bf 1},{\bf 16},0, -\frac{1}4)) + 
\frac12 (k_1 - 1) ({\bf 10},{\bf 1},\os2, 0) +
\nonumber \\
& {} & 
\frac12 (k_2 -1) ({\bf 1},{\bf 10},\os2,0) + 
\frac12 (k_1 + 3)[({\bf 1},{\bf 1},\os2, -\oh) +({\bf 1},{\bf 1},\os2, \oh) ]  +
\nonumber \\
& {} & 4({\bf 1},{\bf 1},\s2,0) 
+ \frac12 (k_2 +3)[({\bf 1},{\bf 1},\os2,-\oh) +  ({\bf 1},{\bf 1},\os2, \oh) ]
+  {\rm c.c.} \big \} +
\nonumber \\
& {} & 
(2k_1+2k_2 + 12)({\bf 1},{\bf 1},0,0) 
\label{rayos}
\eeqa
where we have included the gravitational contribution. 

Since we are setting $m_{1A}=m_{2A}=3$ and $m_{1B}=m_{2B}=2$,
the possible choices for the $SU(2)$ instanton numbers are
$(k_1, k_2)=(7,7), (8,6),$ $(9,5),(10,4),$ and $(11,3)$. It is again
convenient to label the models in terms of the integer
\beq
n\ =\ k_1 + m_{1A} + m_{1B} -   12 
\label{nnuevon}
\eeq
To identify the terminal elements for each $n$ we implement breaking to
$G_0 \times U(1)_Y \times U(1)_Z$, where $G_0$ arises from the second
$SO(10)$ and the surviving $U(1)$'s are the appropriate oblique
combinations of $U(1)_{AD}$, $U(1)_{BD}$ and $SO(10)\times SO(10)$
Cartan generators. Let us briefly discuss the main features of the models 
for the different values of $n$.

\bigskip
\noindent
$n=0,1,2$
\medskip

\noindent
In these cases $(k_1,k_2)=(7,7),(8,6),(9,5)$ respectively.  
The terminal group is $U(1)_Y \times U(1)_Z$ with spectrum consisting of
144 charged and 102 neutral hypermultiplets in all three cases.  
In the 4d Coulomb phase we find a model that matches  
$(b_{21},b_{11})=(101,5)$, in agreement with the Hodge numbers given
in Table~\ref{tabla2}. Un-Higgsing of an $SU(2)$ factor shows different
spectra for different $n$ and gives corresponding Hodge numbers in
accord with (\ref{abchodge}). By Higgsing sequentially 
$U(1)_Y\times U(1)_Z$$\rightarrow U(1)_Y \rightarrow \emptyset$,  
we obtain the corresponding last elements of the $n=0,1,2$ chains
of type B and A.

\bigskip
\noindent
$n=3$
\medskip

Here $(k_1,k_2)=(10,4)$. The terminal group is
$SU(3)\times U(1)_Y\times U(1)_Z$ with 104 charged and 150 
neutral hypermultiplets. In the 4d Coulomb phase we obtain a model 
with corresponding $(b_{21},b_{11})=(103,7)$. These are the Hodge
numbers of $\IP_4^{(1,1,3,5,5)}[15]$.

\bigskip
\noindent
$n=4$
\medskip

In this case $(k_1,k_2)=(11,3)$ and the second $SO(10)$ can be broken
to a matter-free $SO(8)$. Maintaining $U(1)_Y\times U(1)_Z$ also unbroken, 
we arrive at a model with 112 singlet hypermultiplets plus 
extra charged matter. In the 4d Coulomb branch we find 
$(b_{21},b_{11})=(111,9)$. This matches the Hodge numbers of
$\IP_4^{(1,1,4,6,6)}[18]$,  which is the last element of the  
$n=4$ type C chain. Notice that we again have 
an unbroken $SO(8)$ symmetry as it happens in chains of type A and B.

\bigskip
\noindent
$n=5$
\medskip

In this case we would naively set $k_1=12$ and $k_2=2$, but the latter is 
not possible since $k_2=2$ could not support an $SU(2)$ bundle. The 
appropriate distribution of instantons turns out to be
$(k_1,m_{1A},m_{1B};k_2,m_{2A},m_{2B})$ $=(12,3,2;0,3,3)$. Thus, one of 
the two instantons from the removed $SU(2)$ is used to increase 
$m_{2B}$ from 2 to 3 (this is required to avoid inconsistencies in the 
spectrum) whereas the other instanton becomes 
small giving rise to an extra 6d tensor multiplet. 
This is consistent with cancellation of  anomalies, {\it i.e.}
$k_1+m_{1A}+m_{1B}+m_{2A}+m_{2B}+1=24$, where the extra unit comes from
the contribution of the small instanton. The resulting model has
terminal group $E_6 \times U(1)_Y\times U(1)_Z$ plus one tensor multiplet and 
a neutral hypermultiplet. In the 4d Coulomb phase we find 12 vector 
multiplets (one coming from the 6d tensor multiplet) 
and 121 hypermultiplets.  This would correspond to a type IIA  
compactification on a CY with $(b_{21},b_{11})=(120,12)$. These are 
the Hodge numbers of $\IP_4^{(1,1,5,7,7)}[21]$.

\bigskip
\noindent
$n=6$
\medskip

The instanton assignments are $(k_1,m_{1A},m_{1B};k_2,m_{2A},m_{2B})$
$=(13,3,2;0,3,3)$.  The small instanton in the prior situation 
has travelled to the first $E_8$ and acquired a finite size so there are 
no extra tensor multiplets. The terminal group is again 
$E_6\times U(1)_Y \times U(1)_Z$, with 132 singlet hypermultiplets 
plus extra $E_6$ singlets charged under the $U(1)$'s. In the 4d Coulomb branch
we encounter a model with 11 vector multiplets and 132 
hypermultiplets. This corresponds to the Hodge numbers of 
$\IP_4^{(1,1,6,8,8)}[24]$, the last element of the $n=6$  C chain, as expected.

Again, the preceding heterotic construction also predicts the
associated Hodge numbers for many un-Higgsing patterns. For 
$SU(2)\times U(1)^2$ we find results in accord with (\ref{abchodge})
for all $n$. For $SU(3)\times U(1)^2$ and $SU(4)\times U(1)^2$
we obtain
\beqa
b_{12}^4  & = & b_{12}^1 - (8n+25) \quad\quad ; \quad\quad
b_{11}^4  =  b_{11}^1 + 3 \nonumber \\
b_{12}^3  & = & b_{12}^1 - (7n+20) \quad\quad ; \quad\quad
b_{11}^3  =  b_{11}^1 + 2 
\label{csu43}
\eeqa
Other branches emanating from or leading to $SO(10)\times U(1)^2$ can 
be followed as well.

Notice that, in principle, there is no obstruction to the inclusion of more 
Abelian background factors, and therefore to the existence of new types of
models. We only expect to have a shorter range for $n$ since the extra
factors soak up more instanton numbers and, as observed before, 
higher values of $n$ will fall into chains already present.
For instance, D models can be obtained by including $SU(2) \times U(1)^3$ 
backgrounds, with $n= 0,1,2,3,4$ given by the straightforward 
generalization of (\ref{nnuevon}).
As an illustration, notice that in the $n=0$ case, with equal instanton 
numbers in each $SU(2)$ factor,
the initial $SU(5)\times SU(5)$ non-Abelian gauge group  
can be completely 
Higgsed away. This corresponds to a deformation of the initial background to 
$SO(10)\times U(1)^3$ with nine 
instantons in each $SO(10)$. The model obtained has Hodge numbers $(70,6)$.

\subsection{Semisimple Backgrounds}
\label{sub:semi}

Interesting possibilities open when semisimple non-Abelian backgrounds, 
instead of the simple $H$ factors included so far, 
are allowed. In particular, $(R,{\bar R})$ representations, 
leading to higher Kac-Moody level groups with adjoint matter, can
naturally appear. We now want to show how an alternative construction 
for some chains can be achieved in this manner.
We use a notation in which subscripts between parentheses denote instanton
numbers whereas plain subscripts indicate the Kac-Moody level.

As an example, consider an $SU(2)_{(8)} \times SU(2)_{(6)}$ semisimple 
bundle with instanton numbers
$(8,6)$ in the first $E_8$ and an $SU(2)_{(10)}$ bundle with
ten instantons in the second. The observable group is $SO(12)\times E_7$.  
Now we Higgs $E_7$ away and break down $SO(12)$ to $SU(6)$ (which could 
also be obtained by embedding an $SU(2)_{(8)}\times SU(3)_{(6)}$ bundle).
Breaking $SU(6)$ to $SU(5)$ and then continuing along
$SU(4) \rightarrow SU(3) \rightarrow SU(2) \rightarrow \emptyset $, 
we recover the $n=2$ type A chain. This process can be seen as a
deformation of the original $SU(2)\times SU(2)$ through simple group bundles.
Alternatively, one
 can proceed  by maintaining the semisimple structure, {\it e.g.} 
by breaking $SU(6)$ to $SU(3)\times SU(3)$ from the Higgsing point of view.
In this way we arrive at a spectrum
\beq
12[({\bf 3,1})+({\bf {\bar 3},1} ) +({\bf 1,3}) +(\bf {1,{\bar 3}})] + 
({\bf3,\bar 3})+  ({\bf {\bar 3},3}) + 98{\bf (1,1)}
\eeq 
This $SU(3)\times SU(3)$ can also be derived using a
$SU(3)_{(8)}\times SU(3)_{(6)}$ background.
Notice the presence of $ ({\bf3,\bar 3}) + ({\bf {\bar 3},3})$ 
representations that can effect the breaking to the diagonal $SU(3)_2$ 
at level two. Along the direction
$SU(3)_2 \rightarrow SU(2)_2 \rightarrow \emptyset$ we now encounter
matching Hodge numbers $(101,5) \rightarrow (148,4) \rightarrow (243,3)$, 
corresponding to C $\to$ B $\to$ A for $n=2$.
Yet another alternative is to take a different route 
 breaking
$U(1)$ subgroups in each $SU(3)$ to the diagonal combination to arrive 
at $SU(2)\times SU(2)\times U(1)$. Diagonal Higgsing 
then leads to $SU(2) \times U(1)$ and finally $U(1)$. In this way
we reproduce the $n=2$ type B chain 
$(98,6) \rightarrow (121,5) \rightarrow (148,4)$. 
This whole sequence  corresponds to a deformation of the starting 
$SU(2)_{(8)} \times SU(2)_{(6)} $ bundle through
$SU(4)_{(8)} \times SU(3)_{(6)} \rightarrow  SU(7)_{(14)} \rightarrow  
SO(14)_{(14)}$.
Notice again the existence of an $U(1)$ Abelian factor generated here 
through a seemingly different procedure. Finally, let us 
stress that, in spite of this alternative construction for $n=2$,
we have only been able to obtain  a unified
picture for all chains  by considering the $U(1)$ backgrounds
studied in the previous subsections.

The inclusion of semisimple non-Abelian bundles also 
furnishes a possible explanation of the 4d chains mentioned at
the end of last chapter. Let us examine for instance
the chain ending at $(143,7)$.
As discussed, these models are expected to be originated in a
compactification involving enhancing of toroidal $U(1)$'s. We then perform 
a toroidal compactification down to eight dimensions adjusting moduli
parameters in order to obtain an $SO(20)\times E_8$ gauge group as 
in example 10 of ref.~\cite{kv}. The next step is a further 
compactification on $K3$ down to four dimensions, with
an $SU(2)_{(6)}\times SU(2)_{(10)}$ semisimple  bundle 
with $(k_1,k_2)=(6,10)$ instantons embedded in $SO(20)$ and an 
$SU(2)_{(8)}$ bundle with $k_3=8$ in $E_8$. The starting gauge group is 
therefore $SO(12) \times SU(2)\times SU(2)\times E_7$ (not including
dilaton and graviphoton). $E_7$ can be Higgsed down to a 
terminal $SO(8)$ without charged matter. This breaking correponds 
to a deformation of the original $SU(2)_{(8)}$  to an 
$SO(8)_{(8)}$ bundle. Higgsing the other factors is more subtle and we 
proceed as before by systematically breaking to the diagonal $U(1)$ built 
up from Abelian subgroups contained in different factors. 
The last steps correpond to continuous deformations of the original 
bundle according to 
\beq 
SU(3)_{(6)} \times SU(5) _{(10)} \rightarrow SU(8)_{(16)} 
\rightarrow SO(16)_{(16)}
\eeq
Let us check that these bundles do indeed produce the Hodge numbers that
we are looking for. To this purpose we need to compute the dimensions
${\rm dim}\, {\cal M}_k(H)$ of the moduli space associated to each bundle 
$H$ with instanton number $k$. Equivalently, we need to determine the number
of neutral singlets arising from each $H$.  From the index theorem
(\ref{nrabund}) we easily obtain
\beq
{\rm dim}\, {\cal M}_k(H)= k c_H - {\rm dim}\, H
\label{dimmod}
\eeq
where $c_H= T({\rm adj}\, H)$ is the Coxeter number of $H$. Hence,
${\rm dim}\, {\cal M}_8 (SO(8)) =20$, 
${\rm dim}\, {\cal M}_{(6,10)} (SU(3)\times SU(5)) =36$, etc. 
Adding the 20 gravitational moduli we thereby obtain the 4d chain 
of Hodge numbers $(75,9) \rightarrow (104,8)  \rightarrow (143,7)$.

\section{Type II Compactifications}
\label{sec:tipoii}
\bigskip

\subsection{ F-theory Duals of the New Heterotic Models}
\label{sub:ftheory}

Recently a new insight into several string dualities has been provided by
F-theory \cite{vafa}, a construction that can be understood as a type IIB
compactification on a variety B in the presence of  Dirichlet 7-branes. The 
complex `coupling constant' $\tau=a+ie^{-\varphi/2}$, where $a$ is the RR 
scalar field
and $\varphi$ is the dilaton field, depends on space-time and is furthermore
allowed to undergo $SL(2,Z)$ monodromies around the 7-branes. This $\tau$ 
can be thought to describe the complex
structure parameter of a torus (of frozen K\"{a}hler class, since type
IIB theory has no fields to account for it) varying over the
compactifying space $B$, and degenerating at the 8d submanifolds defined
by the world-volumes of the 7-branes. The constraint of having vanishing
total first Chern class (the contribution of the 7-branes cancelling
that of the manifold $B$) forces the $T^2$ over $B$ fibration thus
constructed to be an elliptic CY manifold $X$. 
Thus, F-theory can be understood as a 12d construction
which has consistent compactifications on elliptically fibered manifolds.
It has been conjectured \cite{vafa} that F-theory compactified on the product of 
such an elliptically fibered manifold $X$ and a circle $S^1$, lies on the 
same moduli space as M-theory compactified on $X$. This idea has proved 
fruitful in encoding string dualities in lower dimensions, and, 
especially, in clarifying several phenomena in heterotic string 
compactifications.

After compactification on an elliptic $K3$, F-theory gives an 8d theory
conjectured to be dual to the heterotic string compactified on $T^2$ 
\cite{vafa,sen}. When the elliptic fiber is chosen to be 
$\IP_2^{(1,2,3)}[6]$, the (1,1) $K3$ moduli
and the heterotic moduli are related as follows. The size of
the base $\P1$ is related to the heterotic dilaton whereas the 18
polynomial deformation complex parameters of the fibration match the heterotic 
toroidal K\"{a}hler and complex structure moduli together with Wilson line 
backgrounds. The K\"{a}hler class of the $A_1$ singularity of 
this particular $K3$ 
is associated to the size of the fiber, it has no physical meaning in
F-theory, and thus, no heterotic counterpart.

Fibering this model over another $\P1$ gives a family of F-theory
compactifications on CY 3-folds which are $K3$ fibrations with the $K3$
fibers admitting an elliptic fibration structure. The resulting base
spaces are the Hirzebruch surfaces $\IF_n$, which are fibrations 
of $\P1$ over $\P1$, characterized by a non-negative integer $n$. These 
models are naturally conjectured to be dual to heterotic string 
compactifications on $K3$ ($T^2$ fibered over $\P1$) whith gauge bundles 
embedded on $E_8 \times E_8$ (for some values of $n$, it can also be 
related to $SO(32)$ heterotic string compactifications 
\cite{blpssw, ag}). Upon toroidal compactification to $D=4$, $N=2$ 
heterotic/type II duality is recovered so that
an $N=1$, $D=6$ version of this duality is actually introduced. This has 
several advantages, as dynamics in six dimensions are quite constrained, and 
have been under detailed study in recent works. Concerning this issue, 
let us note that the base space of these compactifications, $\IF_n$, has 
two K\"{a}hler forms, and thus the massless spectum contains only one tensor 
multiplet  (associated to the heterotic dilaton
\cite{mv} ). Consequently, except when 
the singularities in the variety require a blow-up of the base for their 
resolution, we will have $\nt=1$.

Our purpose in this section is to find the F-theory duals of the 
previously discussed heterotic models. We will use the very detailed 
work of ref.~\cite{mv} as a guide. We briefly review some of the main 
results in order to fix the notation and stress the analogies among the 
duals of the different types of heterotic chains.
The choice of specific elliptic fiber $\IP_2^{(1,2,3)}[6]$ implies
CY spaces that can be described as follows. Introducing 
variables $z_1,w_1$ and $z_2,w_2$ for the two $\P1$'s, $x,y$ 
for the torus, and two $\IC^*$ quotients to projectivize the affine spaces,
gives the structure
\beq
\begin{tabular}{ccccccc}
& $z_1$ & $w_1$ & $z_2$ & $w_2$ & $x$ & $y$ \\
$\lambda $ & 1 & 1 & 0 & 0 & 4 & 6  \\
$\mu$ & $n$ & 0 & 1 & 1 & $2n+4$ & $3n+6$ 
\end{tabular}
\label{fiba}
\eeq
The hypersurface in this space is given by the fibration equation
\begin{equation}
y^2 = x^3 + f(z_1,w_1;z_2,w_2) x  + g(z_1,w_1;z_2,w_2)
\label{eq:fibr1}
\end{equation}
It can be shown that for $n>12$ the variety described by (\ref{fiba})
does not fulfill the CY condition (in particular, the
associated Newton polyhedron ceases to be reflexive), so that there are 
13 possible spaces.

For $n \neq 0, 1$, we can dehomogeneize with respect to $w_1$ 
using one of the $\IC^*$ quotients, and the variety can be represented by
the hypersurface $\IP_4^{(1,1,n,2n+4,3n+6)}[6n+12]$. These coincide
with the last elements of the chains of A models. 
Furthermore, for all values of $n$, the Hodge numbers 
of the fibration do match with the
matter spectrum of heterotic models on $K3\times T^2$ with SU(2) bundles of 
$(12+n, 12-n)$
instanton number embedded in $E_8\times E_8$, upon maximal Higgsing and
moving to the Coulomb phase \cite{afiq, cf}. Thus, one identifies type IIA 
compactifications on these spaces with duals of the heterotic 
constructions in $D=4$, or equivalently the F-theory 
compactifications with the heterotic models in $D=6$ (decompactifying the 
$T^2$).

The $ E_8 \times E_8 $ structure in the F-theory compactification 
can be deduced by analyzing the defining equation (\ref{eq:fibr1}) 
near the regions $z_1=0$ and $z_1=\infty$ \cite{mv}. 
We illustrate this strategy since it will be of constant 
reference along this section. To this end we expand the polynomials 
$f,g$ in powers of $z_1,w_1$
\beqa
f(z_1,w_1;z_2,w_2) & = &
\sum_{k=-4}^{4} z_1^{4+k} w_1^{4-k} f_{8-nk}(z_2,w_2)  \nonumber \\
g(z_1,w_1;z_2,w_2) & = &
\sum_{l=-6}^{6} z_1^{6+l} w_1^{6-l} g_{12-nl}(z_2,w_2) 
\label{expan1}
\eeqa
where subscripts 
denote the degree of the polynomial in $z_2$ (only non-negative  
degrees are admitted). 

A first check \cite{mv} of duality consists of the identification of the 
generic type of singular fiber along 
the curves $z_1=0$ and $z_1=\infty$ in the base. The singularity type is 
associated to the terminal gauge group after maximal 
Higgsing in the heterotic side.
Also, some information about the hypermultiplet content of the theory 
can be obtained \cite{mv,bikmsv, katzv}. The existence of hypermultiplets 
charged under the gauge group is detected by the worsening of the 
singularities as $z_2$ varies. It is possible to calculate the numbers of 
singlets coming from each heterotic $E_8$ bundle by counting the number 
of monomial deformations (modulo redundancies) near $z_1=0$ (polynomials of 
type `$f$' in (\ref{expan1}) with $k<0$, and of type `$g$' with $l<0$), 
and near $z_1=\infty$ (polynomials with $k,l>0$). It is interesting to 
note that the outcome of this calculation gives a contribution 
of $[30(12+n)-248]$ and $[30(12-n)-248]$ singlets for the case $n\leq 2$ 
when complete Higgsing of the gauge group is possible. This is a signal
of the already mentioned heterotic $E_8 \times E_8$ bundle structure with 
$(12+n,12-n)$ instantons. The remaining polynomials ($k,l=0$) give the 20 
hypermultiplets associated to the gravitational contribution of the 
heterotic $K3$ compactification. 
A third check  of the equivalence of the heterotic and F-theory description 
of the models comes from the construction of CY spaces dual to heterotic 
models with perturbatively enhanced gauge groups, and with 
non-perturbative effects. In this way, one can reproduce a web of 
F-theory models matching that obtained in the heterotic approach 
\cite{cf,bikmsv}.

The interplay of F-theory and M-theory in the study of 6d heterotic
dynamics has been remarkably useful in checking different ideas about
phase transitions, and new Higgs branches. This motivates the search for
the F-theory duals of the heterotic constructions presented in the 
preceding section. Our analysis for models B, C and D is not as 
detailed as that already performed in the literature for models A, but 
it provides enough evidence to strongly suggest the nature of the 
F-theory duals of the different models and their connections.

The basic idea is to repeat the previous arguments using as elliptic fibers 
other torus realizations such as $\IP_2^{(1,1,2)}[4] $, 
$\IP_2^{(1,1,1)}[3]$, $\IP_3^{(1,1,1,1)}[2,2]$, or even surfaces
in products of projective spaces. These constructions correspond to 
singular Weierstrass models in which the polynomials $f,g$ in 
(\ref{eq:fibr1}) are not as generic as possible, leading to the 
appearance of extra singularities. Consequently,
already in $D=8$, we can see that the resulting fibrations show different 
features compared to those of $\IP_2^{(1,2,3)}[6]$. The $K3$ that they 
give rise to when fibered over $\IP_1$ have less than 18 polynomial 
deformations, the missing moduli being provided by the K\"{a}hler classes 
of the extra $A_1$ singularities. This property leaves its track after 
fibration over another $\IP_1$.

\subsubsection{Type B Models}
\label{sub:typebcy}

We consider the fiber $\IP_2^{(1,1,2)}[4]$ and derive the concrete structure 
of the fibrations as before. The ambient space is defined by 
\beq
\begin{tabular}{ccccccc}
& $z_1$ & $w_1$ & $z_2$ & $w_2$ & $x$ & $y$ \\
$\lambda$ & 1 & 1 & 0 & 0 & 2 & 4 \\
$\mu$ & $n$ & 0 & 1 & 1 & $n+2$ & $2n+4$
\end{tabular}
\label{fibb}
\eeq
and the hypersurface is given by the equation
\begin{equation}
y^2 = x^4 + f(z_1,w_1;z_2,w_2) x^2  + g(z_1,w_1;z_2,w_2) x + 
h(z_1,w_1;z_2,w_2) 
\label{eq:fibr2}
\end{equation}
We also note that, for $n \not= 0,1$, these spaces coincide with
$\IP_4^{(1,1,n,n+2,2n+4)}[4n+8]$. These are precisely the  
last elements of the chains of type B models.
The CY condition for this type of fibration changes, and forces $n$ to be 
between 0 and 8. For this range of $n$, the fibrations have 
Hodge numbers matching our heterotic construction using $SU(2)\times U(1)$
backgrounds with instanton numbers distributed as $(9-n,3;9+n,3)$. 
As we discussed in section 3, these models have an 
enhanced $U(1)$ gauge symmetry, confirming our expectations. Observe that 
the bound on $n$ coincides in both constructions, being associated on the 
heterotic side to the collapse of some bundle structure due to lack of enough 
instantons to support it.

As mentioned earlier, the nature of the heterotic dual can be tested 
by studying the defining equations and the deformation of 
singularities near $z_1=0$ and $z_1=\infty$. The fibration 
defined by (\ref{eq:fibr2}) has an extra singularity responsible for 
the existence of the enhanced $U(1)$ symmetry.   
A detailed analysis of this type of singularity is
still lacking. Nonetheless, we will see that the emergence of  
non-Abelian factors can be correctly deduced. 
Recognizing each particular singularity requires a change of
variables to put eq. (\ref{eq:fibr2}) into Weierstrass form.
The starting point is the
expansion of the polynomial coefficients in (\ref{eq:fibr2}) in 
powers of $z_1,w_1$, namely
 \begin{eqnarray}
f(z_1,w_1;z_2,w_2) & = &
\sum_{i=-2}^{2} z_1^{2+i} w_1^{2-i} f_{4-ni}(z_2,w_2) 
\nonumber \\ 
g(z_1,w_1;z_2,w_2) & = &
\sum_{j=-3}^{3} z_1^{3+j} w_1^{3-j} g_{6-nj}(z_2,w_2) \nonumber \\
h(z_1,w_1;z_2,w_2) & = &
\sum_{k=-4}^{4} z_1^{4+k} w_1^{4-k} h_{8-nk}(z_2,w_2) 
\label{coef2}
\end{eqnarray}
A study of the generic singularities at $z_1=\infty$ (there is no generic 
singularity at $z_1=0$) for each value of 
$n$ reveals agreement with the pattern of heterotic terminal gauge 
groups as we now explain.

For $n=0,1,2$ no singularity is found, as corresponds to having 
complete Higgsing of the gauge group in the heterotic side. 
The $n=3$ case has an $A_2$ singularity, with no $z_2$ dependence, 
so that it maps to an $SU(3)$ gauge group without charged hypermultiplets. 
For $n=4$ we get a $D_4$ singularity, leading to an $SO(8)$ gauge group 
without matter. The $n=5$ case gives an $E_6$ singularity
with quadratic dependence on $z_2$ so that in principle, 
monodromies around the singular fiber could break the 
symmetry to $F_4$ \cite{ag,bikmsv}. However, it can be checked that the 
points at which the singularity worsens pair up to cancel the monodromy and 
the symmetry group $E_6$ remains unbroken with  half 
hypermultiplets in the {\bf 27} representation.
For $n=6$ we find an $E_6$ singularity without $z_2$ dependence that 
maps to a heterotic $E_6$ gauge group with no matter.
The singularity for $n=7$ is of the $E_7$ type, and the linear $z_2$ 
dependence found is associated to a small instanton. Observe that this 
indicates the presence of the $U(1)$ background on the heterotic side 
since, in the absence of instantons, the gauge symmetry is $E_7 (\times 
U(1))$ instead of $E_8$. Finally,
for $n=8$ an $E_7$ symmetry is found, with no small instantons present. 
The same comments as in the precedent case apply here.

The counting of polynomial deformations near $z_1=0$ ({\em i.e.} 
polynomials with $i,j,k<0$) and $z_1=\infty$ ($i,j,k>0$) can be 
performed as before. The number of singlets coming from each bundle is 
found to coincide with heterotic expectations. The result for complete 
Higgsing amounts to $[18(11+n)-133]$ and $[18(11-n)-133]$.
This suggests that the instantons live in an $E_7$ subalgebra of 
$E_8$, a fact that will be relevant for discussions in section 5. 

The main conclusion from the above analysis is the perfect match between 
the F-theory construction and the pattern obtained in class B of  
heterotic compactification ($SU(2)\times U(1)$ bundles).
The analysis for other fibers goes along the same lines. Below we will just 
sketch the main points. 

\subsubsection{Type C Models}
\label{sub:tipoccy}

Fibering $\IP_2^{(1,1,1)}[3]$ over $\IF_n$ leads 
to the defining ambient space
\beq
\begin{tabular}{ccccccc}
& $z_1$ & $w_1$ & $z_2$ & $w_2$ & $x$ & $y$ \\
$\lambda$ & 1 & 1 & 0 & 0 & 2 & 2  \\
$\mu$ & $n$ & 0 & 1 & 1 & $n+2$ & $n+2$ 
\end{tabular}
\label{fibc}
\eeq
and the equation
\begin{eqnarray}
x^3 + y^3 + f(z_1,w_1;z_2,w_2) x y + g(z_1,w_1;z_2,w_2) x + \nonumber\\
g'(z_1,w_1;z_2,w_2) y + h(z_1,w_1;z_2,w_2)=0 
\label{eq:fibr3}
\end{eqnarray}
Again, for $n \not=0,1$, we
find the hypersurfaces $\IP_4^{(1,1,n,n+2,n+2)}[n+6]$ that
give the last elements of the type C chains. 
The Hodge numbers match for all $n$ the
spectrum obtained from heterotic models with $SU(2) \times U(1)^2 $ gauge
backgrounds, as described in section 3. The bound on $n$ due to the CY 
condition is $n\leq6$ in this case, the same found in the heterotic 
construction.

A systematic analysis of the singularities for the different values of 
$n$ leads to a pattern reproducing that of the 
gauge groups of heterotic type C compactifications. We find no 
singularity for $n=0,1,2$. For $n=3$ there is an $A_2$ singularity 
without $z_2$ dependence. For $n=4$ there appears a $D_4$ singularity also
without $z_2$ dependence. For $n=5$ there is an $E_6$ gauge group and  
one small instanton due to the linear $z_2$ dependence of 
the singularity. Lastly, for $n=6$ an $E_6$ singularity without $z_2$ 
dependence is found. A count of parameters perturbing 
the singularities gives also the number of neutral singlets in the model. 
As the conclusion is of some interest, we repeat the exercise explicitly. 
The polynomial deformations can be decomposed as follows
\begin{eqnarray}
f(z_1,w_1;z_2,w_2) & = &
\sum_{i=-1}^{1} z_1^{1+i} w_1^{1-i} f_{2-ni}(z_2,w_2) 
\nonumber \\ 
g(z_1,w_1;z_2,w_2) & = &
\sum_{j=-2}^{2} z_1^{2+j} w_1^{2-j} g_{4-nj}(z_2,w_2) \nonumber \\
g'(z_1,w_1;z_2,w_2) & = &
\sum_{k=-2}^{2} z_1^{2+k} w_1^{2-k} g'_{4-nk}(z_2,w_2) 
\nonumber \\
h(z_1,w_1;z_2,w_2) & = &
\sum_{l=-3}^{3} z_1^{3+l} w_1^{3-l} h_{6-nl}(z_2,w_2) 
\end{eqnarray}
One can check that there exist two extra singularities (so that we 
expect an enhanced $U(1)^2$ gauge symmetry), and that there are 
$[12(10+n)-78]$ parameters deforming the singularity at $z_1=0$ and 
$[12(10-n)-78]$ deforming that at $z_1=\infty$ (in the case of complete 
Higgsing $n\leq 2$). This counting gives the right heterotic result, and 
suggests that the instantons in completely Higgsed models lie on an $E_6$ 
subalgebra of $E_8$. Thus we find enough evidence to support the idea 
that these fibrations provide the duals of heterotic type C models.

\subsubsection{Type D Models}
\label{sub:tipodcy}

We can repeat the construction 
with yet another fiber. The family of CY 3-folds obtained upon fibering 
$\IP_3[2,2] $ over $\IF_n $, is described by
\beq
\begin{tabular}{cccccccc}
& $z_1$ & $w_1$ & $z_2$ & $w_2$ & $x$ & $y$ & $z$ \\
$\lambda $ & 1 & 1 & 0 & 0 & 2 & 2 & 2 \\
$\mu $ & $n$ & 0 & 1 & 1 & $n+2$ & $n+2$ & $n+2$ 
\end{tabular}
\label{fibd}
\eeq
and the pair of equations
\begin{eqnarray}
& {} & x^2 + f(z_1,w_1;z_2,w_2) y + f'(z_1,w_1;z_2,w_2) z +
\nonumber \\ 
& {} & g(z_1,w_1;z_2,w_2) y z + h(z_1,w_1;z_2,w_2)  =  0 \nonumber \\
& {} & y^2 + z^2 + f''(z_1,w_1;z_2,w_2) x + h'(z_1,w_1;z_2,w_2)  =  0
\label{eq:fibr4}
\end{eqnarray}
After eliminating $w_1$ we obtain the CY spaces 
$\IP_5^{(1,1,n,n+2,n+2,n+2)}[2n+4,2n+4]$ for $n\not=0,1$. 
These coincide, for $n=2,4$, with $K3$
fibrations listed in \cite{klm}
and the Hodge numbers coincide with the
spectrum given by heterotic models with $SU(2) \times U(1)^3$ backgrounds 
(the D class of models) that display an enhanced $ U(1)^3 $ gauge group.
A study of the structure of the CY similar to that performed for other 
fibers is also possible in this case. Again the results match those 
obtained from the heterotic type D constructions. We also notice that the 
count of neutral singlets in the case $n\leq 2$ gives 
$[8(9+n)-45]$ and $[8(9-n)-45]$, suggesting a $SO(10)$ structure for the 
instantons after Higgsing.

\subsection{Conifold Transitions}
\label{sub:coni}

We now address the question of the physical process connecting the 
different types of models A, B, C, D. The answer, of course, depends on 
the dimension of space-time, since, as noted in \cite{mv}, in $D=6$ these 
models should not be regarded as different. Since vector multiplets do not 
contain scalars, one cannot turn on vevs to change the K\"{a}hler forms 
(unless they lie on the base $\IF_n$), 
and the vector Coulomb phase is absent. 
Thus, the fibrations with fibers A, B, C, D are related by simply moving 
in the complex structure moduli space to different loci on which the 
Weierstrass models present extra singularities. However, in
$D$=4 this is not the case, and a type IIA string 
compactified on such a singular space can smooth the singularity by 
simply turning on vevs for scalars associated to the K\"{a}her structure 
of the CY, travelling to a new branch of the collective moduli space 
through such a conifold transition \cite{conist, gms}. We now turn to 
working out the details in a concrete example, showing that the CY 
spaces obtained with different fibers and fixed $n$ are connected 
through this process. Also, we note that the transition can be mapped to 
an identical phenomenon in the dual heterotic picture.

As an example we consider the $n=4$ `transversal' chain 
formed with the fibrations of the different elliptic curves over $\IF_4$.
This is
\beqa
& {} & A\, : \, \IP_4^{(1,1,4,12,18)}[36]_{271, 7} \to
B\, : \, \IP_4^{(1,1,4,6,12)}[24]_{164, 8} \to
\nonumber \\
& {} & \hspace{1.0cm} C\, : \, \IP_4^{(1,1,4,6,6)}[18]_{111,9} \to
D \, : \, \IP_5^{(1,1,4,6,6,6)}[12,12]_{76,10}
\label{t4chain}
\eeqa
where subscripts indicate the Hodge numbers. We start with the 
type IIA compactification with (271,7) whose heterotic dual is obtained by 
embedding $SU(2)$ bundles with instanton numbers (16,8) on $E_8\times E_8$. 
The 4d gauge group is $SO(8)\times U(1)^4$ and it 
contains 272 singlets. To study the transition we choose a $\IP_2$ inside 
the CY space, defined by $x_4=x_5=0$. This submanifold will contain all 
the singular points at the conifold locus.
The complex structure of the CY manifold can be adjusted so that only 
monomials containing at least one of the variables $x_4,x_5$ appear in 
the defining equation of the space that can then be written as 
\beq
x_4 g_{24}(x_1,x_2,x_3,x_4,x_5) + x_5 h_{18}(x_1,x_2,x_3,x_4,x_5) = 0
\label{conieq}
\eeq
It then follows that this variety has singularities for the points such that 
$x_4=x_5=g_{24}(x_i)=h_{18}(x_i)=0$. As noted before, all of them live on 
the selected $\IP_2$. The number of singular points is easily computed. 
Equations (\ref{conieq}) have $24\times 18 = 432$ solutions but only
$108=432/4$ are distinct due to the scaling symmetries of the projective 
space. For generic choices of the polynomials 
$g_{24}, h_{18}$, the singularities have the typical conical structure with 
base $S^3\times S^2$. The specific choice of complex structure has moved us 
to a conifold locus of dimension 164 (the number of independent
polynomial deformations not smoothing the singularities) and codimension 
107 (the number of vanishing 3-cycles minus one homology relation among 
them).

Observe that the tuning of the complex structure is performed through vevs 
for the hypermultiplets. Hence, it maps to an un-Higgsing of the $U(1)$ gauge 
group on the heterotic side. It is illustrative to consider the spectrum 
of this heterotic model, listed in equation (\ref{ncuatro}). The $U(1)$ 
gauge boson is associated to the K\"ahler class of the 
2-cycle associated to the small resolution of the 
singularities. The 165 
neutral singlets on the heterotic side correspond to the 164 type IIA 
fields (plus the dilaton) that can receive vevs without 
destroying the singularity (heterotic $U(1)$), while 108 fields can move 
us out of the singular locus (charged fields Higgsing the $U(1)$). 

The singularities can be resolved by increasing the size of the $S^2$'s. 
Along this branch, a new K\"{a}hler class appears, and the 107 complex 
structure deformations associated to the 3-cycles are lost. This realizes 
the dual mechanism of the heterotic Higgsing with the 
scalar in the $U(1)$ vector multiplet, in which charged fields become 
massive.
As the number of vector and hypermultiplets has changed, we 
have landed in the moduli space of another Calabi-Yau, in our case of 
Hodge numbers $(271-107,7+1)=(164,8)$. This spectrum and the precise 
mapping with the heterotic process 
suggest that we have arrived at the space $\IP_4^{(1,1,4,6,12)}[24]$.

It is also possible to show that the complex structure of model B in
(\ref{t4chain}) can be adjusted to develop conifold singularities 
whose resolution leads to the Hodge numbers of model C. The transition from
C to D is similarly verified. The whole `transversal' chain is 
also reproduced for other values of $n$
\footnote{The $C \to D$ transition for $n=2$ was noticed in ref.~\cite{ls}.}
and is expected to be correct 
in general, even though some transitions may be harder to check.
For example, in some cases $b_{11}$ increases in several units, reflecting 
more homology relations among the 3-cycles.
The interesting point in this discussion is to notice how all the 
connections in the vast web of models that we have explored can be studied 
from both the heterotic and the type IIA points of view, thus leading to 
a clearer picture of the way duality acts on these constructions.

\section{ Phase Transitions }

\bigskip

The $E_8 \times E_8$ compactifications obtained with $H\times U(1)^{8-d}$
backgrounds in $K3$ all have the same kind of strong coupling singularity
found in the case $d=8$. For example, for the $B$ models with group
$E_6 \times E_6 \times U(1)_D$, the anomaly polynomial (\ref{ane6e6}) 
implies gauge kinetic terms of the form
\beq
-\frac13 (e^{-\phi}+ \frac{n}2e^{\phi})\tr\, F_1^2 -
 \frac13 (e^{-\phi}-\frac{n}2e^{\phi})\tr\, F_2^2 -
 (e^{-\phi} + 3 e^{\phi}) f^2
\label{bsing}
\eeq
where we have set $k_1=9+n$ and $k_2=9-n$. Thus, the strong coupling
singularity occurs again at the dilaton value (\ref{phtr}).
For models of type C and D we also find that the coupling of respectively
the second $SO(10)$ and $SU(5)$ diverges at (\ref{phtr}).
We now discuss several features of the singularities following the
analysis of refs. \cite{sw, mv, witphase}. 

The singularity is signaled by the appearance of tensionless strings
\cite{gh, sw, dlp} that in the F-theory approach arise from a threebrane
wrapping around a vanishing rational curve in the base \cite{mv}.
Since we are dealing with elliptic fibrations over $\IF_n$, general 
arguments \cite{mv} imply that when $n=2$ there is actually no
singularity since the collapsed $\IF_2$ can be deformed into $\IF_0$
and the gauge coupling is not singular when $n=0$. 
This particular behaviour of the $n=2$ case can  be
understood from the heterotic point of view as arising
from the fact that, in that case,  one can completely 
Higgs away the gauge group which is related to the singularity
 \cite{afiq2}. For $n=1,4$,
general arguments also indicate that at the singularity there 
occurs a transition to a Higgs branch with no dilaton. We now review the 
supporting evidence of this type of transition for the new class of
models discussed in the previous sections.

When $n=1$, the transition is described by a change of base from $\IF_1$
to $\IP_2$. Counting the change in parameters agrees with the results  
expected from a heterotic process in which the dilaton tensor
multiplet disappears. For example,
for models B, the fibration over $\IF_1$ has an equation of the form
(\ref{eq:fibr2}) with coefficients $f,g$ and $h$ of bidegrees (4,6), (6,9)
and (8,12) in $\lambda$ and $\mu$. The total number of monomials 
in $(z_1,w_1,z_2,w_2)$ is 155
but there are 7 redundant parameters \cite{witphase} due to transformations
among the variables. Altogether there are 148 independent parameters.
On the other hand, the fibration over $\IP_2$ has equation
\beq
y^2 = x^4 + \tilde f(z_1,z_2,w_2) x^2  + \tilde g(z_1,z_2,w_2) x +
\tilde h(z_1,z_2,w_2) 
\label{basep2}
\eeq
where now $\tilde f, \tilde g$ and $\tilde h$ are homogeneous functions
of degrees 6,9 and 12 in $\lambda$. There are 174 possible monomials in
$(z_1,z_2,w_2)$ and 9 redundant parameters so that the number of 
independent deformations is 165. 

The above counting exercise basically amounts to a determination of
the Hodge number $b_{12}$ of the fibration over $\IF_1$ and the fibration
over $\IP_2$ that happens to be $\IP_4^{(1,1,1,3,6)}[12]$.
The difference $\Delta(b_{12}) = 165-148=17$ can be explained in the 
heterotic side by considering instanton numbers (10,3;9,3), and one
less tensor multiplet, compared to the $n=1$ instantons (10,3;8,3).
The values of $k_1$ and $k_2$ are such that $E_6 \times E_6$ can be
completely broken leaving behind the spectrum
\beq
\{ 54(\frac1{\sqrt2}) + {\rm c.c.} \}
+ 166(0)
\label{n1new}
\eeq
where we have taken into account the fact that when $\nt=0$ there is
one less neutral hypermultiplet. Notice that compared to (\ref{ncud})
we do have $29$ extra hypermultiplets as required by (\ref{gravan}). 
However, twelve of them are charged and become massive in the Coulomb phase.

A similar analysis applies to models C in which the fibration of
$\IP_2[3]$ over $\IP_2$ is $\IP_4^{(1,1,1,3,3)}[9]$. In 
this case we find $\Delta(b_{12}) = 112-101=11$, in agreement with the 
heterotic spectrum obtained for instanton numbers (8,3,2;7,3,2) and $\nt=0$.
For models D the fibration of $\IP_3[2,2]$ over $\IP_2$ is the 
hypersurface  $\IP_5^{(1,1,1,3,3,3)}[6,6]$ and we find 
$\Delta(b_{12}) = 77-70=7$. Again this is in agreement with results
obtained in the heterotic side.

When $\nt$ decreases by one, the number of 4d vector multiplets 
is lowered by one. Correspondingly, $b_{11}$ decreases in one unit since 
there is one collapsing 2-cycle. Thus, in all models we can write
\beqa
\Delta(b_{11}) & = & -1 \nonumber \\
\Delta(b_{12}) & = & c_d-1
\label{coxeter}
\eeqa
where $c_d$ is the Coxeter number of $E_d$ and $d=8,7,6$ and 5 for models
A,B,C and D. The groups $E_d$ do enter in the heterotic picture as
follows. Notice that for $n=1$, complete Higgsing of the non-Abelian
groups is possible in all models and this can be achieved by instantons of 
$E_d \times U(1)^{8-d}$ that leave $U(1)^{8-d}$ unbroken in each $E_8$
(before further breaking to the diagonal combinations). In fact, the 
transition to $n_T=0$ occurs when $k_2 \to k_2 + 1$, where $k_2$ corresponds 
to an $E_d$ instanton. In the F-theory picture, the $E_d$ groups appear
because when the 2-cycle collapses in $\IF_1 \to \IP_2$, there also
shrinks a 4-cycle of del Pezzo type \cite {mv}. 
In turn, this del Pezzo surface is related
to the form of the singularity at $w_1=0$ \cite{mv}. For example, for
models C with $d=6$, from (\ref{eq:fibr3}) we see that, setting say $z_1=1$,
the singularity is locally a hypersurface in $\IC^4$ with leading
cubic terms. Similarly, for models $D$ with $d=5$, (\ref{eq:fibr4})
implies that the singularity is locally the intersection of two
quadratic equations in $\IC^5$. 

It is also possible to probe the current algebra carried by the 
tensionless string that develops when an instanton shrinks \cite{gh}
in the reverse transition $\IP_2 \to \IF_1$. In F-theory a rank one
current algebra is supported at the intersection of a type IIB 3-brane
and a type IIB 7-brane \cite{witphase}. The idea is then to determine
the number of 7-branes that meet the 2-cycle blanketed by the 3-brane.
In turn this can be done by counting the parameters of the fibration
restricted to $w_1=0$ \cite{vafa, witphase}. For example, for models
$B$, from eq.~(\ref{coef2}) we find 3 parameters in $f$, 4 in $g$ and 5
in $h$. Eliminating the redundancies due to linear transformations of
$(z_2,w_2)$ leaves 8 independent parameters. This indicates then that
the 3-brane intersects eight 7-branes so that the current algebra has 
rank eight. The same result readily follows for models C and D. 

Existence of a Higgs branch with zero tensor multiplets is also expected in
the strong coupling transition for the $n=4$ case, on the basis of anomaly
cancellation arguments \cite{sw} and F-theory computations
\cite{mv}. Since, in the latter approach, the transition corresponds to a
deformation of the base of the fibration from $\IF _4$ to $\IP _2$, it
follows that such kind of transitions will be possible not only for A,
but also for B, C and D models. Actually, it is evident that the change 
in the Hodge numbers for the associated CY spaces follows the rule
\beqa
\Delta(b_{11}) & = & -4 \nonumber \\
\Delta(b_{12}) & = & 1
\label{coxetern4}
\eeqa
which can be understood as follows. As the tensor multiplet disappears,
anomaly cancellation conditions force the appearance
of 29 new hypermultiplets, 28 of them are employed in Higgsing the $SO(8)$
gauge symmetry and one remains in the final spectrum providing for the
increase in $b_{21}$. In this process
4 Cartan generators are lost, thus explaining the change in $b_{11}$. 
It is important to notice the relevance of the $SO(8)$
symmetry for this counting to work (on the F-theory side, the existence
of the corresponding $D_4$ singularity is discussed in \cite {mv}). 
As remarked in previous sections, this requirement
is fulfilled by all $n=4$ models. An interesting point in this discussion
is that the new hypermultiplets appearing in the transition seem to be 
charged under the terminal gauge group $SO(8)$. Also, in the new branch 
there is no generic gauge symmetry as corresponds to the F-theory fibration 
over $\IP _2$. The mechanism of smoothing the singularity  
gives a hint about how this occurs. As described in \cite{mv}, it is
related to the $Z_2$ quotient of the deformation of 
$\IP _2 \to \IP _1 \times \IP _1$. Since in this process the instanton 
numbers embedded in each $E_8$ change from a $(14,10)$ to a $(12,12)$ 
distribution, we expect a similar change in the initial $(16,8)$
instanton distribution for $n=4$. In this way the bundle in the
second $E_8$ ends up with enough instantons to achieve complete
Higgsing of the gauge group.

A comment concerning  transitions
to $D=6$ models with no tensor multiplets is in order.
Just looking at the spectrum, 
from a purely 4d point of view, it may be difficult to
disentangle whether the dual of a certain type II CY 
compactification
is a perturbative ($n_T=1$)  or a non-perturbative 
($n_T=0$) heterotic vacuum. Let us consider the $n=4$ case. 
For the A chain this can be
obtained by embedding instanton numbers 
$(k_1,k_2)=(16,8)$ in $E_8\times E_8$. Higgsing as much
as possible the second $E_8$ we arrive at a $D=6$ model
with gauge group $E_7\times SO(8)$ and hypermultiplets
transforming as $6({\bf {56}}, {\bf 1})+69({\bf 1},{\bf 1})$.
In addition there is a $D=6$ tensor multiplet containing the
dilaton. Now consider the  model obtained by embedding instantons
with $(k_1,k_2)=(16,9)$ in $E_8\times E_8$. This is the final
stage of a model in which the original tensor multiplet has
been absorbed at the M-theory boundary and has been
converted into an instanton in the second $E_8$. 
Thus this model is continuously connected to the previous
one. Since
there is no dilaton to make a perturbative expansion this
is a non-perturbative vacuum. We can Higgs in steps 
the second $E_7$ of this theory. If we stop at an
$SO(10)$ stage, the gauge group will be $E_7\times SO(10)$
and it is easy to check that there will be hypermultiplets 
transforming as  
$6({\bf {56}}, {\bf 1})+69({\bf 1},{\bf 1})$
$+({\bf 1}, {\bf {16}})+3({\bf 1},{\bf {10}})$.
Now, the point is that if we further compactify these two 
models on $T^2$, Higgs completely the first $E_7$  
and go to the Coulomb phase, we arrive in both 
cases to a $N=2$ model with the same number of vector multiplets
and hypermultiplets, corresponding to a type II compactification
on a CY with $(b_{21},b_{11})=(271,7)$.
For the perturbative heterotic vacuum the seven vector 
multiplets correspond to $7=S+T+U+{\rm rank}\, (SO(8))$  whereas 
for the non-perturbative model one has
$7=T+U+{\rm rank}\, (SO(10))$, in an obvious notation. 
We know that these
two models are connected through a transition
$n_T=1\rightarrow n_T=0$. Hence, one can argue that 
the $SO(8)$ group of the first model 
can combine with the dilaton vector multiplet  
to get a non-perturbatively enhanced  $SO(10)$.

\section{Final Comments and Conclusions}
\label{sec:fin}

In the heterotic constructions of section~\ref{sec:hetk3}, Abelian  
backgrounds played an essential role. They provide a systematics for 
deriving chains of different types, each type corresponding to the 
inclusion of a given number of $U(1)$'s. For a given set of instanton 
numbers, specified by $n$, many Higgsing branches can be followed. 
Continuous flow from an $n$ fixed branch of a given type to another type 
is achieved by Higgsing $U(1)$ gauge groups. These processes have a
dual description in terms of transitions in the space of CY spaces.
It must be emphasized that the full web of dual theories is quite
intricate. Identifying precisely special points such as the terminal
A,B,C and D models provides a useful handle in exploring this web.

We have shown that the process of changing the fiber 
in F-theory compactifications is associated to the appearance of 
enhanced gauge symmetries arising from $E_8\times E_8$.
As one can embed a larger number of $U(1)$ backgrounds in $E_8$ on the 
heterotic side, we expect to find further families of CY
spaces associated to other F-theory fibers. 
Also, once the last elements of the chains have been understood, all models
corresponding to un-Higgsing of  gauge symmetries
should be derivable using the techniques  presented in 
refs.~\cite{cf,bikmsv}, leading to an extended web of models on the F-theory 
side.
It would be interesting, for instance, to study type B and C
duals with enhanced $SU(r)$ groups and compare their Hodge numbers
with those implied by the heterotic analysis of section 3.

The heterotic models discussed to large extent, all
arise from $E_8\times E_8$ compactifications. However, in some situations,
there appear suggestive correlations when $SO(32)$ compactifications
are examined. For instance, it is well known that starting with the 
standard embedding in $SO(32)$ leads, for  generic moduli, to a model with 
Hodge numbers $(271,7)$ and a matter-free terminal $SO(8)$, the same
result found in the terminal $n=4$ A model. In fact, the equivalence 
between both constructions was established in \cite{blpssw} by using 
T-duality arguments. Moreover, the same authors show that the symmetric
instanton embedding $(12,12)$, {\it i.e.} the $n=0$ type A case
is equivalent to an $SO(32)$ compactification without vector structure. 
This corresponds to the Type I string model elaborated in \cite {gp}.
We have found extra examples
that suggest additional relations between $SO(32)$ and $E_8\times E_8$
compactifications.

The first example is a six-dimensional $Z_3$ orbifold compactification
accompanied by the embedding of the shift 
$V= \frac{1}{3} (-2,1,1,1,1,1,1,1,1,1,1,0,0,0,0,0)$ in the $Spin(32)/Z_2$
lattice. The resulting model has gauge group
$SU(11)\times SO(10)  \times U(1)$ and massless hypermultiplet spectrum given by
\begin{eqnarray}
\theta ^0:& &  ({\bf {11}}, {\bf {10}},-1)+ ({\bf {55}}, {\bf {1}},-2)
+2({\bf {1}}, {\bf {1}},0) \nonumber \\
\theta ^1:& &  9[({\bf {11}}, {\bf {1}},{\frac {2}3 })+
({\bf {1}}, {\bf {1}},{\frac {5}3 }) +
({\bf {\bar {10}}}, {\bf {1}}, {-\frac {4}3 })]
\label{f1}
\end{eqnarray}
The gauge group can be completely Higgsed away, leading to $(243,3)$  Hodge
numbers. Moreover, first breaking $SO(10)$ fully and then
performing a cascade breaking of $SU(11)$, the chain
\begin{equation}
\dots \rightarrow(193,8) \rightarrow (204,6)
\rightarrow (215,5) \rightarrow (226,4)
\rightarrow (243,3)
\label{f2}
\end{equation}
is obtained.
This is  the same set of numbers found for $n=1$ type A models
with instanton numbers $(13,11)$, if the first $E_8$
(with $k_1=13$) is completely Higgsed and the second is broken sequentially.
Interestingly enough, an alternative Higgs branch can be followed through
level two models with adjoints, due
to the presence of the $({\bf {11}}, {\bf {10}})$ representation.
The Hodge numbers
$(70,6) \rightarrow(101,5) \rightarrow (148,4) \rightarrow (243,3)$
are derived in this way. This corresponds to transversal transitions 
D $\to$  C $\to$ B $\to$ A among the $n=1$ terminal elements.
 
Constructions in terms of semisimple bundles in $SO(32)$ are also 
interesting. For instance, by embedding an 
$SU(8)_{(k_1)}\times SU(8)_{(k_2)}$ bundle in 
$SO(32)$ it is easy to see that for instanton numbers $(k_1,k_2)=(12,12), 
(13,11), (14,10)$, full Higgsing is possible ending in the $(243,3)$ model.
Another interesting example starts with background
$SU(2)_{(4)}\times SU(2)_{(6)}\times SU(2)_{(14)}$ in $SO(32)$ to
obtain observable group 
$SU(2)^3\times SO(20)$. Higgsing through steps similar to those 
discussed in Chapter 3, the  bundle may be deformed to 
$SU(2)_{(4)}\times SU(3)_{(6)}\times SO(14)_{(14)} \rightarrow  
SU(3)_{(6)}\times SO(18)_{(18)} \rightarrow  SO(24)_{(24)}$. 
Using eq.~(\ref{dimmod}) to compute the number of moduli we encounter the 
sequence $(111,9) \to (164,8) \to (271,7)$, corresponding to 
transitions through last elements of different types for $n=4$

In conclusion, in this paper we have studied 
new branches of $D=6,4$ heterotic string compactifications
obtained by including Abelian backgrounds 
on the $E_8\times E_8$ heterotic string. The corresponding
type II duals can be derived from F-theory by changing
appropriately the elliptic fiber. Our procedure allows us to
explicitly construct the heterotic duals of  many
type IIA   compactifications on $K3$ fibrations whose duals
were previously unknown. It also allows us to understand 
the existence of  some chains of models which were
conjectured to be connected in  \cite{afiq} . 
The connections between the different types of
chains of models are understood in terms of the
Higgsing of $U(1)$'s in the heterotic side and
conifold transitions from the type II side.
We also identify new  $D=6$ models  in which transitions from 
theories with one  tensor multiplet to zero tensor 
multiplets occur.
Other interesting features appear in our class of models.
In particular, there are anomalous $U(1)$'s that are
in fact Higgsed away by swallowing zero modes
of the antisymmetric $B_{MN}$ field. A similar phenomenon
was recently reported in \cite{blpssw}. 

Although most of the work reported here is related to compactifications
of the $E_8\times E_8$ heterotic string, it is clear that related
models may be obtained from $SO(32)$. It would be interesting to study 
also these models and their connections to type II compactifications.

\vskip0.6cm

\centerline{\bf Acknowledgments}
\bigskip

We thank F. Quevedo, who participated in the early stages of this
work, for useful discussions.
We enjoyed enlightening conversations with L.~Alvarez-Gaum\'e, P.~Candelas,
E.~Derrick, R.~Hern\'andez, D.~L\"ust and M.~Mandelberg.
We are grateful to P.~Candelas and A.~Klemm for verification of Hodge
numbers. G.A. thanks the Departamento de F\'{\i}sica Te\'orica
at UAM for hospitality, and the Ministry of Education and Science of Spain
as well as CONICET (Argentina) for financial support.
A.F. thanks CONICIT (Venezuela) for a research grant S1-2700
and CDCH-UCV for a sabbatical fellowship. A.M.U. thanks the Governement
of the Basque Country for financial support.

\newpage
\begin{table}
{\scriptsize
\begin{tabular}{|c|c|c|c|c|c|c|} 
\hline
$n$ & \multicolumn{2}{c|}{A} & \multicolumn{2}{c|}{B} &
\multicolumn{2}{c|}{C} \\[0.2ex]
\cline{2-7}
{} & $(b_{12},b_{11})$ & Weights & $(b_{12},b_{11})$ & Weights & 
$(b_{12},b_{11})$ & Weights \\[0.2ex]
\hline
\hline
2  & (138,6) & (1,1,2,6,8,10)  &  &  &  & \\
{} &(161,5) & (1,1,2,6,8) & (102,6) & (1,1,2,4,6,6) &  &  \\
{} & (190,4) & (1,1,2,6,10) & (121,5) & (1,1,2,4,6)  & (82,6) & 
(1,1,2,4,4,6) \\
{} &  (243,3) & (1,1,2,8,12)  & (148,4) & (1,1,2,4,8) & (101,5) & 
(1,1,2,4,4) \\ 
\hline 
3  & (124,8) &  (1,1,3,7,9,11)  &  &  &  & \\
{} & (151,7) & (1,1,3,7,9)  & (96,8) & (1,1,3,5,7,9) &  & \\
{} & (186,6) & (1,1,3,7,12) & (119,7) & (1,1,3,5,7) & (80,8) & (1,1,3,5,5,7) \\ 
{} & (251,5) & (1,1,3,10,15) & (152,6) & (1,1,3,5,10) & (103,7) &
(1,1,3,5,5) \\
\hline
4  & (122,10) & (1,1,4,8,10,12) &  &  &  & \\
{} & (153,9) & (1,1,4,8,10) & (98,10) & (1,1,4,6,8,10) &  & \\
{} & (194,8) & (1,1,4,8,14) & (125,9) & (1,1,4,6,8) & (84,10) &
(1,1,4,6,6,8) \\
{} & (271,7) & (1,1,4,12,18) & (164,8) & (1,1,4,6,12) & (111,9) & 
(1,1,4,6,6) \\
\hline
5  & (124,10) & (1,1,5,9,11,13) &  &  &  & \\
{} & (159,9) & (1,1,5,9,11) & (102,12) & (1,1,5,7,9,11) &  & \\
{} & (206,8) & (1,1,5,9,16)  & (133,11) & (1,1,5,7,9) & (89,13) & 
(1,1,5,7,7,9) \\
{} & (295,7) & (1,1,5,14,21) & (178,10) & (1,1,5,7,14) & (120,12) & 
(1,1,5,7,7)\\
\hline
6  & (128,12) & (1,1,6,10,14) &  &  &  & \\
{} & (167,11) & (1,1,6,10,12) & (108,12) & (1,1,6,8,10,12) &  & \\
{} & (220,10) & (1,1,6,10,18) & (143,11) & (1,1,6,8,10) & (96,12) & 
(1,1,6,8,8,10) \\
{} & (321,9)  & (1,1,6,16,24) & (194,10) & (1,1,6,8,16) & (131,11) &
(1,1,6,8,8) \\
\hline
7  & (133,13) & (1,1,7,11,13,15) &  &  &  & \\
{} & (176,12) & (1,1,7,11,13) & (114,14) & (1,1,7,9,11,13)  &  & \\
{} & (235,11) & (1,1,7,11,20) & (153,13) & (1,1,7,9,11) &  &  \\
{} & (348,10) & (1,1,7,18,27) & (210,12) & (1,1,7,9,18)  &  & \\
\cline{1-5}
8  & (139,13) & (1,1,8,12,14,16) &  &  &  & \\
{} & (186,12) & (1,1,8,12,14) & (121,13) &  (1,1,8,10,12,14) &  & \\
{} & (251,11) & (1,1,8,12,22) & (164,12) & (1,1,8,10,12) &  &  \\
{} & (376,10) & (1,1,8,20,30) & (227,11) & (1,1,8,10,20) &  & \\
\cline{1-5}
9  & (145,17) & (1,1,9,13,15,17) &  &  &  & \\
{} & (196,16) & (1,1,9,13,15) &  &  &  &   \\
{} & (267,15) & (1,1,9,13,24)  &  &  &  & \\
{} & (404,14) & (1,1,9,22,33) &  &  &  & \\
\cline{1-3}
10  & (152,16) & (1,1,10,14,16,18) &  &  &  & \\
{} & (207,15) & (1,1,10,14,16) &  &  &  &   \\
{} & (284,14) & (1,1,10,14,26)  &  &  &  & \\
{} & (433,13) & (1,1,10,24,36) &  &  &  & \\
\cline{1-3}
11 & (159,15) & (1,1,11,15,17,19) &  &  &  & \\
{} & (218,14) & (1,1,11,15,17)  &  &  &  & \\
{} & (301,13)  & (1,1,11,15,28) &  &  &  & \\
{} & (462,12) & (1,1,11,26,39) &  &  &  & \\
\cline{1-3}
12 & (166,14) & (1,1,12,16,18,20) &  &  &  & \\
{} & (229,13) & (1,1,12,16,18) &  &  &  & \\
{} & (318,12) & (1,1,12,16,30) &  &  &  & \\
{} & (491,11) & (1,1,12,28,42) &  &  &  & \\
\hline
\end{tabular}
\caption{Hodge numbers for the chains of CY spaces.}
\label{tabla3}
}
\end{table}

\end{document}